\documentclass[%
reprint,
nofootinbib,
amsmath,amssymb,
aps,
]{revtex4-2}

\usepackage{graphicx}
\usepackage{dcolumn}
\usepackage{bm}
\usepackage[figtopcap]{subfigure}
\usepackage{multirow}
\usepackage[colorlinks,citecolor=blue,urlcolor=blue,linkcolor=blue]{hyperref}

\begin{document}

\preprint{APS/123-QED}

\title{Phenomenological reconstruction of $f(T)$ teleparallel gravity}

\author{W.~El~Hanafy}
 \email{waleed.elhanafy@bue.edu.eg}
 \altaffiliation[Also at ]{Egyptian Relativity Group, Cairo University, Giza 12613, Egypt.}
\author{G.~G.~L.~Nashed}%
 \email{nashed@bue.edu.eg}
\affiliation{%
 Centre for Theoretical Physics, the British University in Egypt, El Sherouk City 11837, Egypt.
}%


\begin{abstract}
We present a novel reconstruction method of $f(T)$ teleparallel gravity from phenomenological parametrizations of the deceleration parameter or other alternatives. This can be used as a toolkit to produce viable modified gravity scenarios directly related to cosmological observations. We test two parametrizations of the deceleration parameter considered in recent literatures in addition to one parametrization of the effective (total) equation of state (EoS) of the universe. We use the asymptotic behavior of the matter density parameter as an extra constraint to identify the viable range of the model parameters. One of the tested models shows how tiny modification can produce viable cosmic scenarios quantitatively similar to $\Lambda$CDM but qualitatively different whereas the dark energy (DE) sector becomes dynamical and fully explained by modified gravity not by a cosmological constant.
\end{abstract}

\maketitle


\section{Introduction}\label{S1}
Cosmological observations have confirmed that the universe has speeded up its expansion rate few billion years ago. This includes (i) the age of the universe compared to oldest stars, (ii) supernovae observations \cite{Riess:1998cb,Perlmutter:1998np}, (iii) cosmic microwave background (CMB) \cite{Aghanim:2018eyx}, (iv) baryon acoustic oscillations (BAO) \cite{2011MNRAS.416.3017B,Ross:2014qpa,Alam:2016hwk,Ata:2017dya,2018MNRAS.tmp.3191A}, and (v) large-scale structure (LSS) \cite{Colless:2001gk}. In return, this requires us either to introduce an exotic matter species with negative EoS, the so-called DE, or to consider the repulsive side of gravity by modifying the general relativity theory. DE in its simplest version is represented by a cosmological constant with EoS $w=-1$, that is $\Lambda$CDM model. In general, one needs field equations
\begin{equation}\label{MG}
    \frac{1}{\kappa^2_{eff}}\mathfrak{G}_{\mu\nu}=\mathfrak{T}_{\mu\nu}+\mathfrak{T}^{DE}_{\mu\nu},
\end{equation}
where the effective gravitational function $\kappa^2_{eff}\to \kappa^2(=8\pi G)$ when GR is restored, the Einstein tensor $\mathfrak{G}_{\mu\nu}$ and the matter-stress tensor $\mathfrak{T}_{\mu\nu}$. However, the DE contributes via the tensor $\mathfrak{T}^{DE}_{\mu\nu}$. There are two main approaches to feed the DE component $\mathfrak{T}^{DE}_{\mu\nu}$: (i) physical DE, (ii) geometrical DE \cite{Sahni:2006pa}. Although the two approaches are qualitatively different, the field equations in form (\ref{MG}) allows us to quantitatively treat both of them in a similar way.

Since successful cosmological descriptions should perform deceleration--to--acceleration at a late phase compatible with observations, it is reasonable to code the late accelerated expansion phase via the deceleration parameter $q(z)$. A kinematical approach has been adopted for that purpose by suggesting parametrizations of the deceleration parameter in the form of $q(z)=q_0+q_1 X(z)$, where the two model parameters $q_0$ and $q_1$ are fixed by observations \cite{Riess:2004nr,Cunha:2008ja,Cunha:2008mt,Santos:2010gp,Nair:2011tg,Mamon:2016dlv,Mamon:2017rri, Shafieloo:2007cs,Holsclaw:2010nb,Holsclaw:2011wi,Crittenden:2011aa,Nair:2013sna,Sahni:2006pa}. However, this approach does not provide an explanation of the nature of the DE. The aim of this paper is to set this kinematical approach within a modified gravity theory which should also allow for further tests on the perturbation level of the theory.

We organize the paper as follows. In Sec. \ref{Sec2}, we give a brief account of the $f(T)$ teleparallel gravity and the corresponding modifications of Friedmann equations in cosmological applications. In Sec. \ref{Sec3}, we show the compatibility of the $f(T)$ teleparallel gravity field equations with the deceleration parameter which allows for a nice reconstruction method of $f(T)$ gravity. In this method, one can obtain the $f(T)$ gravity which generates a particular parametric form of the deceleration parameter. In addition, we derive two reconstruction equations of $f(T)$ gravity by knowing the effective EoS $w_{eff}(z)$ or the DE EoS $w_{DE}(z)$. In Sec. \ref{Sec4}, we derive the $f(T)$ gravity which produces the $\Lambda$CDM model using our reconstruction method. In addition, we examine two parametrizations of the deceleration parameter \cite{Mamon:2015osa,Mamon:2016dlv} within the corresponding $f(T)$ theories. Furthermore, we examine one parametric form of the effective EoS \cite{Mukherjee:2016eqj}. We also discuss the results of each of the three models. Finally, we summarize the paper in Sec. \ref{Sec5}.
\section{$f(T)$ Teleparallel Gravity of FLRW universe}\label{Sec2}
We begin this section by a brief introduction to the teleparallel geometry. Let $(M,\,h_{a})$ be a space, where $M$ is a $4$-dimensional smooth manifold equipped with $4$--independent vector fields (tetrad) defined globally on $M$ at each point, $h_{a}$ ($a=1,\cdots, 4$). The vectors $h_a$ satisfy the orthonormality $h_{a}{^{\mu}}h^{a}{_{\nu}}=\delta^{\mu}_{\nu}$ and $h_{a}{^{\mu}}h^{b}{_{\mu}}=\delta^{b}_{a}$, where ($\mu = 1, \cdots, 4$) are the coordinate components of the vector $h_{a}$. Notably, the tetrad fields satisfy the absolute parallelism condition $\nabla_{\nu}h_{a}{^{\mu}}\equiv 0$, whereas the differential operator $\nabla_{\nu}$ is the covariant derivative associated with the Weitzenb\"{o}ck connection $\Gamma^{\alpha}{_{\mu\nu}}\equiv h_{a}{^{\alpha}}\partial_{\nu}h^{a}{_{\mu}}=-h^{a}{_{\mu}}\partial_{\nu}h_{a}{^{\alpha}}$. Several applications have been developed within this geometrical framework, c.f. \cite{Wanas:1986zz,Nashed:2003ee,Wanas:2012ve,Hehl:1994ue}, for more detail see \cite{Youssef:2006sz,Aldrovandi:2013wha}. Interestingly, this connection has a vanishing curvature tensor, however it defines the torsion tensor $T^\alpha{_{\mu\nu}}\equiv{\Gamma^\alpha}_{\nu\mu}-{\Gamma^\alpha}_{\mu\nu}={h_a}^\alpha\left(\partial_\mu{h^a}_\nu
-\partial_\nu{h^a}_\mu\right)$. Consequently, the contortion tensor can be given as $K_{\alpha \mu \nu}=\frac{1}{2}\left(T_{\nu\alpha\mu}+T_{\alpha\mu\nu}-T_{\mu\alpha\nu}\right)$. It is important to remember that the tetrad fields define a metric tensor on $M$ via $g_{\mu \nu} \equiv \eta_{ab}h^{a}{_{\mu}}h^{b}{_{\nu}}$ with an induced Minkowskian metric $\eta_{ab}$ on the tangent space, where the inverse metric is given as $g^{\mu \nu} = \eta^{ab}h_{a}{^{\mu}}h_{b}{^{\nu}}$. In this sense, one can reconstruct the Levi-Civita connection on $M$, and then the Riemannian geometry can be performed.

In teleparallel geometry, one can define the scalar $T$, which is known as the teleparallel torsion scalar. This is given by
\begin{equation}
T \equiv {T^\alpha}_{\mu \nu}{S_\alpha}^{\mu \nu},\label{Tor_sc}
\end{equation}
where the superpotential tensor
\begin{equation}
{S_\alpha}^{\mu\nu}=\frac{1}{2}\left({K^{\mu\nu}}_\alpha+\delta^\mu_\alpha{T^{\beta\nu}}_\beta-\delta^\nu_\alpha{T^{\beta \mu}}_\beta\right),\label{superpotential}
\end{equation}
is skew symmetric in the last pair of indices. Since the teleparallel torsion scalar, $T$, differs from the Ricci scalar $R$ by an additive total derivative term, the resulting field equations are just equivalent to the general relativity when $T$ is employed as a Lagrangian instead of $R$ in Einstein-Hilbert action. So it is a teleparallel equivalent version of the general relativity (TEGR) theory of gravity.
\subsection{FLRW spacetime}\label{Sec2.1}
We take the flat Friedmann-Lema\^{\i}tre-Robertson-Walker (FLRW) metric,
\begin{equation}\label{FRW-metric}
ds^2=dt^{2}-a(t)^{2}\delta_{ij} dx^{i} dx^{j},
\end{equation}
where $a(t)$ is the scale factor of the universe. The above metric can be reconstructed via diagonal vierbein
\begin{equation}\label{tetrad}
{h_{\mu}}^{a}=\textmd{diag}\left(1,a(t),a(t),a(t)\right).
\end{equation}
Interestingly, the teleparallel torsion scalar (\ref{Tor_sc}) of the FLRW spacetime is related directly to Hubble parameter by
\begin{equation}\label{TorHubble}
T=-6H^2,
\end{equation}
where $H\equiv \dot{a}/a$ is Hubble parameter where the dot denotes the derivative with respect to the cosmic time $t$.
\subsection{Field equations}\label{Sec2.2}
We take the action of a matter field minimally coupled to gravity
\begin{equation}\label{action}
    \mathcal{S}=\int d^{4}x |h|\left(\mathcal{L}_{g}+\mathcal{L}_{m}\right),
\end{equation}
where $|h|=\sqrt{-g}=\det\left({h}_\mu{^a}\right)$, $\mathcal{L}_{g}$ and $\mathcal{L}_{m}$ are the Lagrangians of gravity and matter, respectively. Inspired by the $f(R)$-gravity which replaces $R$ by an arbitrary function $f(R)$ in the Einstein-Hilbert action, the TEGR has been generalized by replacing $T$ by an arbitrary function $f(T)$ \cite{BF09, L10,1008.4036,1011.0508}. In the natural units ($c=\hbar=k_{B}=1$), the $f(T)$ Lagrangian is
\begin{equation}\label{gravity-Lag}
    \mathcal{L}_{g}=\frac{1}{2\kappa^2}\,f(T).
\end{equation}
Then, the variation of the action (\ref{action}) with respect to the tetrad fields gives rise to the set of the field equations (\ref{MG}). In the framework of the $f(T)$ modified gravity, we write
\begin{eqnarray}
\mathfrak{T}{_{\mu}}{^{\nu}}&=&h^{a}{_\mu}\left(-\frac{1}{h}\frac{\delta \mathcal{L}_{m}}{\delta h^{a}{_\nu}}\right), \label{Tmn}\\
\kappa^2_{eff}&=&\frac{\kappa^2}{f_T},\label{eff_coupl}\\
\mathfrak{T}^{DE}_{\mu\nu}&=&\frac{1}{\kappa^2} \left(\frac{1}{2}g_{\mu\nu}\left(Tf_T-f\right)-f_{TT}S_{\nu\mu\rho}\nabla^{\rho}T\right),\label{Tmn_DE}
\end{eqnarray}
where $f_T=df/dT$ and $f_{TT}=d^2f/dT^2$. By setting $f(T)=T$, the general relativistic limit is recovered, where $\mathfrak{T}^{DE}_{\mu\nu}$ vanishes and $\kappa_{eff}\to \kappa$. Interestingly enough, this form allows to deal with the torsional and the physical DE on equal footing. We note that the teleparallel torsion scalar is not local Lorentz invariant, which directly leads to the conclusion that the field equations of the nonlinear $f(T)$ are not invariant under local Lorentz transformation \cite{Li:2010cg,Sotiriou:2010mv}. However, a later invariant version of the $f(T)$ gravity has been obtained by considering the contribution of the spin connection to the field equations \cite{Krssak:2015oua}, see also \cite{Jarv:2019ctf}. The $f(T)$ teleparallel gravity has been considered essentially in the recent literature as an alternative to inflation at early universe \cite{Ferraro:2006jd,Bamba:2016gbu,Awad:2017ign} as well as an alternative to dark energy at late universe \cite{BF09, L10,1008.4036,1011.0508,Myrzakulov:2010vz}. Also it has been used to describe bounce cosmology \cite{Cai:2011tc,deHaro:2014tla,Haro:2015zta,ElHanafy:2017sih}. For more reading about $f(T)$ teleparallel gravity, see \cite{Cai:2015emx,Nojiri:2017ncd}.

We assume the stress-energy tensor to be for perfect fluid as
\begin{equation}\label{matter}
 \mathfrak{T}{_{\mu\nu}}=\rho_{m} u_{\mu}u_{\nu}+p_{m}(u_{\mu}u_{\nu}+g_{\mu\nu}),
\end{equation}
where $u^{\mu}$ is the fluid 4-velocity unit vector. Inserting the vierbein (\ref{tetrad}) into the field equations (\ref{MG}), and by making use of the useful relation (\ref{TorHubble}), the $f(T)$ version of Friedmann equations can be given as
\begin{eqnarray}
  \rho_{m} &=& ~\frac{1}{2\kappa^2}\left(f-H f_{H}\right), \label{FR1H}\\
\nonumber  p_{m}    &=& -\frac{1}{2\kappa^2}\left(f-H f_{H}-\frac{1}{3}\dot{H} f_{H}\right)\\
&=&\frac{1}{6\kappa^2}\dot{H} f_{HH}-\rho_{m}, \label{FR2H}
\end{eqnarray}
where $f_H=df/dH$ and $f_{HH}=d^2f/dH^2$.
\subsection{Dynamical view of the modified Friedmann equations}\label{Sec2.3}
In order to close the system, one should choose an EoS to relate $\rho_{m}$ and $p_{m}$. For the simplest barotropic case $p_{m}\equiv p_{m}(\rho_{m})=w_{m}\rho_{m}$, the above system produces the useful dynamical equation
\begin{equation}\label{phase-portrait}
    \dot{H}=3(1+w_{m})\left[\frac{f(H)-H f_{H}}{f_{HH}}\right]=\mathcal{F}(H).
\end{equation}
Since $\dot{H}$ as clear from the above relation is a function of $H$ only, then Eq. (\ref{phase-portrait}) defines the phase portrait of an arbitrary $f(T)$ gravity for flat FLRW background. In the rest of the paper, we focus our analysis on the late cosmic evolution, thence we assume that the universe is dominated by the baryons matter $w_{m}=0$ in the last phase before transition to DE domination.

Since the general relativity shows amazing results with observations, modified gravity theories should be recognized as corrections of it. So it is always useful to rewrite the field equations in away showing Einstein's gravity in addition to the higher order $f(T)$ teleparallel gravity as correction terms. Thus, we write the modified Friedmann equations as
\begin{eqnarray}
{H}^2= \frac{\kappa^2}{3} \left( \rho_{m}+  \rho_{ T} \right)&\equiv& \frac{\kappa^2}{3} \rho_{eff}, \label{MFR1}\\
2 \dot{{H}} + 3{H}^2= - \kappa^2 p_{ T }&\equiv& -\kappa^2 p_{eff}.\label{MFR2}
\end{eqnarray}
In this case, the density and pressure of the torsional counterpart of $f(T)$ are defined by
\begin{eqnarray}
\rho_{T}(H)&=&\frac{1}{2\kappa^2}\left(H f_{H}-f(H)+6H^{2}\right),\label{rhoT}\\
p_{T}(H)&=&-\frac{1}{6\kappa^2}\dot{H}\left(12+f_{HH}\right)-\rho_{T}(H).\label{pT}
\end{eqnarray}
At the GR limit ($f(T)=T$), we have $\rho_{T}=0$ and $p_{T}=0$. For nonlinear $f(T)$ cases, the torsional counterpart of $f(T)$ could play the role of the DE. In the barotropic case, the torsion will have an EoS
\begin{equation}\label{torsion_EoS}
   w_{DE}=w _{T}(H)=-1-\frac{1}{3}\frac{\dot{H}(12+f_{HH})}{6H^2-f(H)+Hf_{H}}.
\end{equation}
Introducing the density parameters $\Omega_{i}=\rho_{i}/\rho_{c}$, where the label $i$ indicates the species component and $\rho_{c}$ is the critical density ($\equiv \rho_{eff}$). Thus, the dimensionless form of Friedmann equation (\ref{MFR1}) is written as
\begin{equation}\label{dimless_FR}
    \Omega_m+\Omega_T=1,
\end{equation}
where $\Omega_m=\frac{\kappa^2\rho_m}{3H^2}$ the matter density parameter and $\Omega_T=\frac{\kappa^2\rho_T}{3H^2}$ is the torsion density parameter. To fulfill the conservation principle, when the matter field and the torsion are minimally coupled, we have the continuity equations
\begin{eqnarray}
  \dot{\rho}_{m}+3H \rho_{m} &=& 0, \label{sc-continuity}\\
  \dot{\rho}_{T}+3H(\rho_{T}+p_{T}) &=& 0. \label{tor-continuity}
\end{eqnarray}
It is useful also to define the effective EoS parameter
\begin{equation}\label{eff_EoS}
w _{eff}\equiv \frac{p_{eff}}{\rho_{eff}}=-1-\frac{2}{3}\frac{\dot{H}}{H^2}.
\end{equation}
The effective EoS parameter can be considered as an alternative to the deceleration parameter $q$, since they are related by
\begin{equation}\label{deceleration}
    q\equiv -1-\frac{\dot{H}}{H^{2}}=\frac{1}{2}\left(1+3w_{eff}\right).
\end{equation}
It is confirmed by several cosmological observations that the universe has turned its expansion from deceleration to acceleration few billion years ago. Since then the deceleration parameter $q$ has been widely used to describe the cosmic history at least at that phase until now. In this sense, some used different parametric form of $q$, cf. \cite{Riess:2004nr,Cunha:2008ja,Cunha:2008mt,Santos:2010gp,Nair:2011tg,Mamon:2016dlv,Mamon:2017rri}, others used nonparametric forms of $q$, c.f. \cite{Shafieloo:2007cs,Holsclaw:2010nb,Holsclaw:2011wi,Crittenden:2011aa,Nair:2013sna,Sahni:2006pa}. However, these needs to be formulated within a framework of a gravitational theory. In the next section, we show how to reconstruct $f(T)$ gravity from a given form of $q(z)$, where $z$ is the redshift, or other parameters as $w_{eff}(z)$. This allows us to perform extra tests on the free parameters of these forms using other cosmological parameters like the matter density parameter $\Omega_m$ or the DE EoS.
\section{Reconstruction Method of $f(T)$ Teleparallel Gravity}\label{Sec3}
In this section we are interested to reconstruct the $f(T)$ gravity upon parametric forms of the deceleration $q(z)$ or other alternatives. So it is convenient to use the redshift $z$ as independent variable, where $z=\frac{a_{0}}{a}-1$ with $a_{0}=1$ at the present time. In this case, we write
\begin{equation}\label{redsh-Hubble}
\dot{H}=-(1+z)H H',
\end{equation}
where the prime denotes the derivative with respect the redshift parameter $z$. Using (\ref{deceleration}) and (\ref{redsh-Hubble}), we write
\begin{equation}\label{Hubble-deceleration}
    H(z)=H_{0}\exp\left(\int_{0}^{z}\frac{1+q(\bar{z})}{1+\bar{z}}d\bar{z}\right),
\end{equation}
where $H_{0}=H(z=0)$. On the other hand, by using $z$ as independent variable, we have
\begin{equation}\label{z-trans}
    f(H(z))=f(z), \, f_{H}=\frac{f'}{H'}, \, f_{HH}=\frac{f'' H'- f' H''}{H'^{3}}.
\end{equation}
By substituting from (\ref{redsh-Hubble}) and (\ref{z-trans}) in the $f(T)$ phase portrait (\ref{phase-portrait}), we evaluate
\begin{equation}\label{phase-portrait-z}
    H(z)=H_{0}\exp\left(\int_{0}^{z}\frac{df/d\bar{z}}{f(\bar{z})+f_{0}(1+\bar{z})^{3}} d\bar{z}\right),
\end{equation}
where $f_{0}$ is a constant of integration. We determine the evolution of the density of the matter component, by substituting from (\ref{redsh-Hubble}) and (\ref{z-trans}) in (\ref{FR1H}), we write
\begin{equation}\label{matter-density}
    \rho_{m}(z)=\frac{1}{2\kappa^2}\left(f(z)-\frac{H}{H'}f'\right)=-\frac{f_{0}}{2\kappa^2}(1+z)^{3}.
\end{equation}
On the other, one can determine the matter density by solving the matter continuity (\ref{sc-continuity}) whereas $\rho_m=\rho_{m,0}a^{-3}=\rho_{m,0}(1+z)^3$ and $\rho_{m,0}$ is the current matter density. By comparison with (\ref{matter-density}), we determine that $f_0=-2\kappa^2\rho_{m,0}$. At the present time (i.e $z=0$), we have $\Omega_{m,0}=\rho_{m,0}/\rho_{c,0}=\frac{\kappa^2 \rho_{m,0}}{3 H_0^2}$, this gives
\begin{equation}\label{f0}
f_{0}=-6\Omega_{m,0}H_{0}^{2}.
\end{equation}
It is worthwhile to mention that the Planck CMB measures the quantity $\Omega_{m,0}h^2=0.1426 \pm 0.0020$ (based on the $\Lambda$CDM model fitted to Planck TT+lowP likelihood), where $h=H_0/100$ km/s/Mpc \cite{Ade:2015xua}. This directly fixes the value of the constant $f_{0}=-8556$.

Obviously, the deceleration parameter $q(z)$ and the modified $f(T)$ gravity are both related to Hubble function $H(z)$ very similarly as indicated by Eqs (\ref{Hubble-deceleration}) and (\ref{phase-portrait-z}). So, by comparing these, we obtain
\begin{equation}\label{deceleration-f(z)}
q(z)=-1+\frac{(1+z)f'}{f(z)-6\Omega_{m,0}H_0^2(1+z)^{3}}.
\end{equation}
As noted before that the deceleration parameter directly reflects the nature of the cosmic expansion rate. For this reason, several parametrization forms of the decelerations have been suggested in the literature, c.f. \cite{Riess:2004nr,Cunha:2008ja,Cunha:2008mt,Santos:2010gp,Nair:2011tg,Mamon:2016dlv,Mamon:2017rri, Shafieloo:2007cs,Holsclaw:2010nb,Holsclaw:2011wi,Crittenden:2011aa,Nair:2013sna,Sahni:2006pa}, to describe the cosmic evolution. Thus, we find that equation (\ref{deceleration-f(z)}) is a toolkit to develop a viable cosmic scenarios within an $f(T)$ gravitational theory.

Notably, the deceleration parameter, (\ref{deceleration}), contains only up to the first derivative of the Hubble parameter, $\dot{H}$. On the other hand, we clarify that the modified version of Friedmann equations in the case of $f(T)$ teleparallel gravity can be rewritten as a one dimensional autonomous system, i.e. $\dot{H}=\mathcal{F}(H)$, see Eq. (\ref{phase-portrait}). This feature cannot be found in modified gravity with a curvature base, e.g. the $f(R)$ gravity, since the dependence of the higher derivative terms, $\ddot{H}$, should violate this feature. In this sense, we find that the Hubble parameter can be expressed as integrals of $q(z)$ and $f(T(z))$ in similar ways, as seen from Eqs. (\ref{Hubble-deceleration}) and (\ref{phase-portrait-z}). The comparison of these two expressions allows the reconstruction method of the $f(T)$ gravity as presented in this paper. Therefore, this approach are not expected to be carried out for $f(R)$ gravity.

Alternatively, Eq. (\ref{deceleration-f(z)}) can be integrated to the construction equation
\begin{equation}\label{Reconstruction1}
    f(z)=-6\Omega_{m,0}H_0^2e^{{\textstyle{\int_0^z}}\frac{1+q(\bar{z})}{1+\bar{z}}d\bar{z}}{\displaystyle{\int_0^z}}
    \frac{(1+\bar{z})^{2}(1+q(\bar{z}))}{e^{{\textstyle{\int_0^z}}\frac{1+q(\bar{z})}{1+\bar{z}}d\bar{z}}}d\bar{z}.
\end{equation}
In the above we have omitted an additional term proportional to the quantity $e^{{\textstyle{\int_0^z}}\frac{1+q(\bar{z})}{1+\bar{z}}d\bar{z}}$, since it represents a total derivative term $H\propto \sqrt{-T}$ in the action (\ref{action}) and does not contribute in the field equations. Thus, for a given parametrization of $q(z)$, Eq. (\ref{Reconstruction1}) enables to generate the corresponding $f(T)$ theory, and then other important parameters can be computed and confronted with observational results to examine the validity of the $f(T)$ gravity. In the present paper, we use the reconstruction equation (\ref{Reconstruction1}) to evaluate the $f(T)$ gravitational theory which generates a proposed parametrization $q(z)$. Also we may interchange $q(z)$ and $w_{eff}(z)$ as given by (\ref{deceleration}), and then $f(T)$ gravity can be reconstructed as well for different parametrizations of $w_{eff}(z)$. In this case, we have another reconstruction equation form
\begin{eqnarray}
\nonumber   f(z)&=&-9\Omega_{m,0}H_0^2e^{\frac{3}{2}{\textstyle{\int_0^z}\frac{1+w_{eff}(\bar{z})}{1+\bar{z}}}d\bar{z}} \\
&&\times\int_0^z
    \frac{(1+\bar{z})^{2}(1+w_{eff}(\bar{z}))}
    {e^{\frac{3}{2}{\textstyle{\int_0^z \frac{1+w_{eff}(\bar{z})}{1+\bar{z}}}} d\bar{z}}} d\bar{z}.\label{Reconstruction2}
\end{eqnarray}
In order to cover other possible reconstruction methods, we include the case when some parametrizations are given for the EoS of the DE sector. Inserting (\ref{redsh-Hubble}) and (\ref{z-trans}) into (\ref{torsion_EoS}), we write
\begin{eqnarray}
w_{T}(z)=-1+\frac{1}{3}(1+z)H\frac{12H'^{3}+f''H'-f'H''}{(6H^{2}-f)H'^{2}+HH'f'}.\label{wT(z)}
\end{eqnarray}
The above equation can be used to reconstruct $f(z)$ from a given parametrization of the DE EoS $w_T(z)$. This can be done by inserting (\ref{phase-portrait-z}) into (\ref{wT(z)}), then we have a new reconstruction equation
\begin{widetext}
\begin{equation}\label{Reconstruction3}
    w_T(z)=\frac{\left[f(z)-6\Omega_{m,0}H_0^2(1+z)^3-\frac{2}{3}(1+z)\frac{df}{dz}\right]
    e^{2{\textstyle{\int_0^z}}\frac{df/d\bar{z}}{f(\bar{z})-6\Omega_{m,0}H_0^2(1+\bar{z})^3} d\bar{z}}}
    {\left[f(z)-6\Omega_{m,0}H_0^2(1+z)^3\right]\left[\Omega_{m,0}(1+z)^3-e^{2{\textstyle{\int_0^z}}\frac{df/d\bar{z}}{f(\bar{z})-6\Omega_{m,0}H_0^2(1+\bar{z})^3}d\bar{z}}\right]}.
\end{equation}
\end{widetext}
Interestingly enough by inserting Eqs. (\ref{Hubble-deceleration}) and (\ref{Reconstruction1}) into (\ref{wT(z)}), we obtain a useful relation between the DE EoS and the deceleration parameter
\begin{equation}\label{wT-deceleration}
    w_{T}(z)  =\frac{\left(1-2q(z)\right)e^{2{\textstyle{\int_0^z}} \frac{1+q(\bar{z})}{1+\bar{z}} d\bar{z}}}{3\left(\Omega_{m,0}(1+z)^3- e^{2{\textstyle{\int_0^z}} \frac{1+q(\bar{z})}{1+\bar{z}} d\bar{z}}\right)},
\end{equation}
Again we may use $q(z)$ and $w_{eff}(z)$ interchangeably via (\ref{deceleration}), then we also relate the DE and the total (effective) EoS parameters as follows
\begin{equation}\label{wT-weff}
    w_{T}(z)  =-\frac{w_{eff}(z)e^{3{\textstyle{\int_0^z} \frac{1+w_{eff}(\bar{z})}{1+\bar{z}} d\bar{z}}}}{\Omega_{m,0}(1+z)^3- e^{3{\textstyle{\int_0^z} \frac{1+w_{eff}(\bar{z})}{1+\bar{z}} d\bar{z}}}},
\end{equation}

In order to close this section, we relate the deceleration parameter to another fundamental cosmological parameter which allows for testing assumed parametrization forms of $q(z)$, that is the matter density parameter $\Omega_{m}(z)$. Using (\ref{matter-density}), we write
\begin{equation}\label{matter-density-parameter}
\Omega_{m}(z)=\frac{fH'-Hf'}{6H' H^{2}}=\Omega_{m,0}(1+z)^{3}\, e^{-2\textstyle{\int_0^z \frac{1+q(\bar{z})}{1+\bar{z}}d\bar{z}}}.
\end{equation}
Also, the dark torsional counterpart is then given by
\begin{equation}\label{torsion-density-parameter}
\Omega_{T}=1-\Omega_{m}=1-\Omega_{m,0}(1+z)^{3}\, e^{-2\textstyle{\int_0^z \frac{1+q(\bar{z})}{1+\bar{z}}d\bar{z}}}.
\end{equation}
We summarize this section by emphasizing on the three reconstruction equations (\ref{Reconstruction1}), (\ref{Reconstruction2}) and (\ref{Reconstruction3}) that allow to reconstruct the $f(T)$ gravity for a given parametric form of the deceleration, the effective EoS and the DE EoS parameters, respectively. We also provide two useful supplementary equations, namely (\ref{wT-deceleration}) and (\ref{wT-weff}), which relate the dark torsional EoS to the deceleration and the effective EoS parameters, respectively. On the other hand, the matter density parameter, Eq. (\ref{matter-density-parameter}), and its asymptotic behavior provides one more test to examine the validity of the suggested parametrization. In the following section, we use these equations in order to examine some parametric forms within $f(T)$ gravity.
\section{Models}\label{Sec4}
Motivated by the results of Sec. \ref{Sec3}, we present three different parametrizations, two for the deceleration parameter $q(z)$ and one for the effective EoS parameter $w_{eff}(z)$, aiming to construct the corresponding $f(T)$ gravity and test possible viable deviations from the $\Lambda$CDM model.
\subsection{Flat $\Lambda$CDM model}\label{Sec4.1}
One of the approaches to to describe the late accelerated expansion phase is DE scenario. The simplest version is to relate it to a cosmological constant, where the cosmic expansion can transform from deceleration to acceleration in a way very compatible with wide range of observations, that is $\Lambda$CDM cosmology. However, the model lakes theoretical interpretations and justifications. In this model, the Hubble evolution is given as
\begin{equation}\label{HLCDM}
    H(z)=H_0\sqrt{\Omega_{m,0}(1+z)^3+\Omega_{\Lambda,0}} \, ,
\end{equation}
where $\Omega_{\Lambda,0}=1-\Omega_{m,0}$ denotes the present value of the DE density parameter. Substituting from (\ref{HLCDM}) into (\ref{deceleration}) taking into account (\ref{redsh-Hubble}), we write the $\Lambda$CDM deceleration parameter
\begin{equation}\label{qLCDM}
    q(z)=-1+\frac{3}{2}\frac{\Omega_{m,0}(1+z)^3}{\Omega_{\Lambda,0}+\Omega_{m,0}(1+z)^3}.
\end{equation}
As clear, at large redshifts, $(1+z)^3 \gg \frac{\Omega_{\Lambda,0}}{\Omega_{m,0}}$, the model gives a decelerated expansion phase in agreement with Einstein-de Sitter model, i.e. $q\to 1/2$. However, at low redshifts, the expansion goes to accelerated phase as $q(z)$ becomes negative, then it evolves toward pure de Sitter $q\to -1$ as $z\to -1$ ($t\to\infty$). Thus, this form gives a viable cosmological scenario in agreement with observations. In addition, we can find the corresponding $f(T)$ gravity representation by by inserting (\ref{qLCDM}) in the reconstruction equation (\ref{Reconstruction1}),
\begin{equation}\label{fLCDM}
    f(z)=-6H_0^2\left(\Omega_{m,0}(1+z)^3+\Omega_{\Lambda,0}\right)-6\Omega_{\Lambda,0}H_0^2.
\end{equation}
This turns out as expected to $f(T)_{\Lambda CDM}=T-const.$, i.e. $\Lambda$CDM model \cite{Nesseris:2013jea}. So the model has two parameters, $H_0$ and $\Omega_{m,0}$, need to be constrained by observations. In fact, the local CMB and BAO observations (preassume $\Lambda$CDM model) favor a value of $H_0\simeq 68$ km/s/Mpc and $\Omega_{m,0}\simeq 0.3$, this is on contrary to the global SNIa and $H_0$ observations (model independent) which favor a larger $H_0\simeq 73$ km/s/Mpc and smaller $\Omega_{m,0}\simeq 0.26$. However, several approaches have been suggested to to reconcile local with global measurements by introducing dark neutrino species \cite{DiValentino:2015ola}, using dynamical phantom DE \cite{DiValentino:2016hlg,Zhao:2017cud} or by utilizing infrared gravity \cite{El-Zant:2018bsc}, see also \cite{Nunes:2018xbm,Escamilla-Rivera:2019ulu}. In modified gravity framework one may search for an explanation for the accelerated expansion without assuming DE (cosmological constant), however we should not expect great deviations from the $\Lambda$CDM scenario.
\subsection{Model 1}\label{Sec4.2}
Several parametrization forms of the deceleration parameter have been suggested in the literature, but in general, they have the following form
\begin{equation}\label{general-param-q}
    q(z)=q_{0}+q_{1}X(z),
\end{equation}
where $q_{0}$ and $q_{1}$ are real numbers can be fixed by observational datasets, while different choices of the function $X(z)$ give different parametrizations of the deceleration parameter. Motivated by the Barboza and Alcaniz parametrization of the DE EoS \cite{Barboza:2008rh}, a divergence-free parametrization of the deceleration
parameter has been suggested as \cite{Mamon:2015osa}
\begin{equation}\label{Mod1-Xz}
    X(z)=\frac{z(1+z)}{1+z^2}.
\end{equation}
\begin{figure*}[t!]
\centering
\subfigure[~$f(T)$ gravity evolution]{\label{fig:Mod1-fz}\includegraphics[scale=0.3]{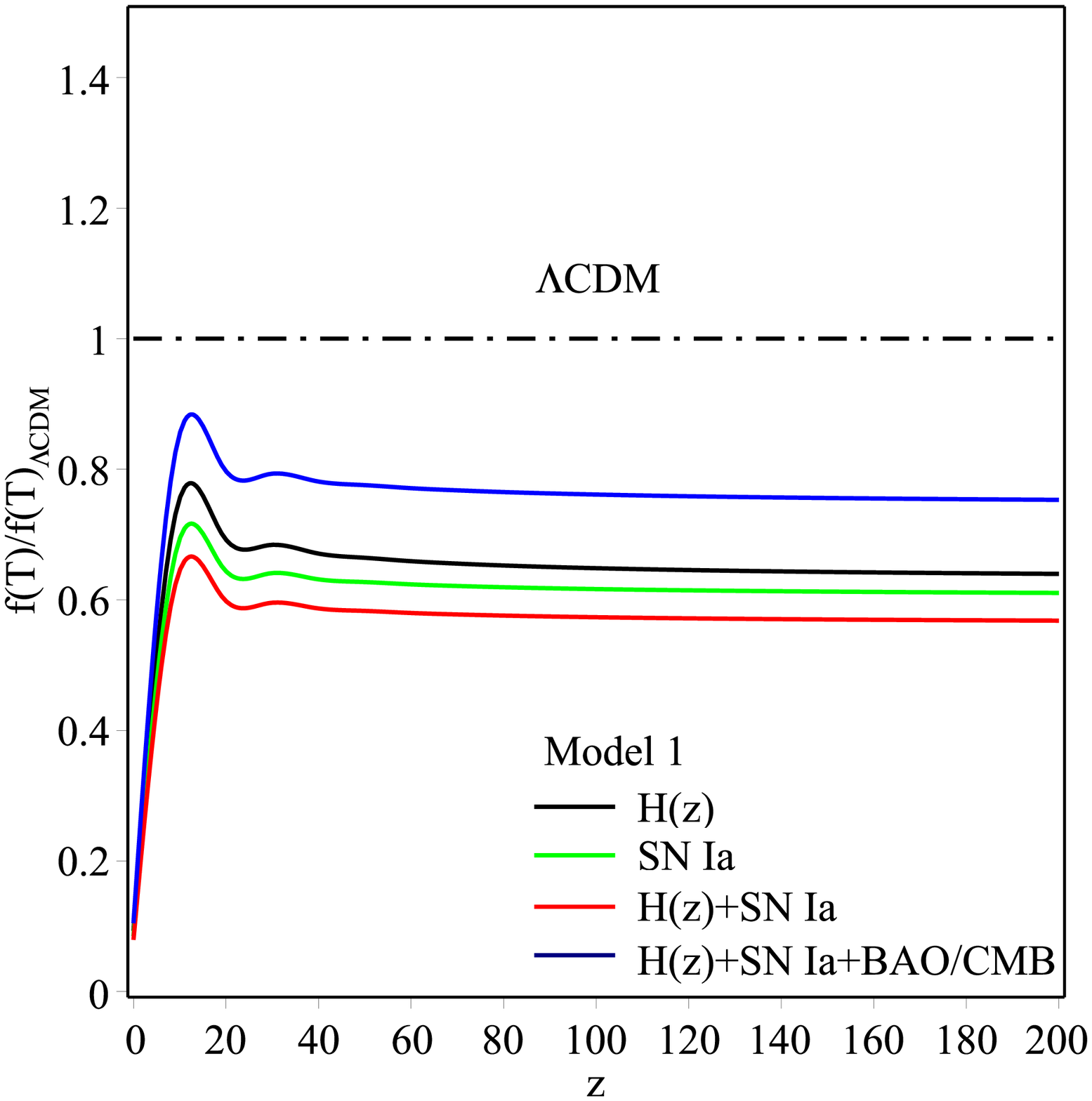}}
\subfigure[~$w_{eff}$ evolution]{\label{fig:Mod1-weff}\includegraphics[scale=.3]{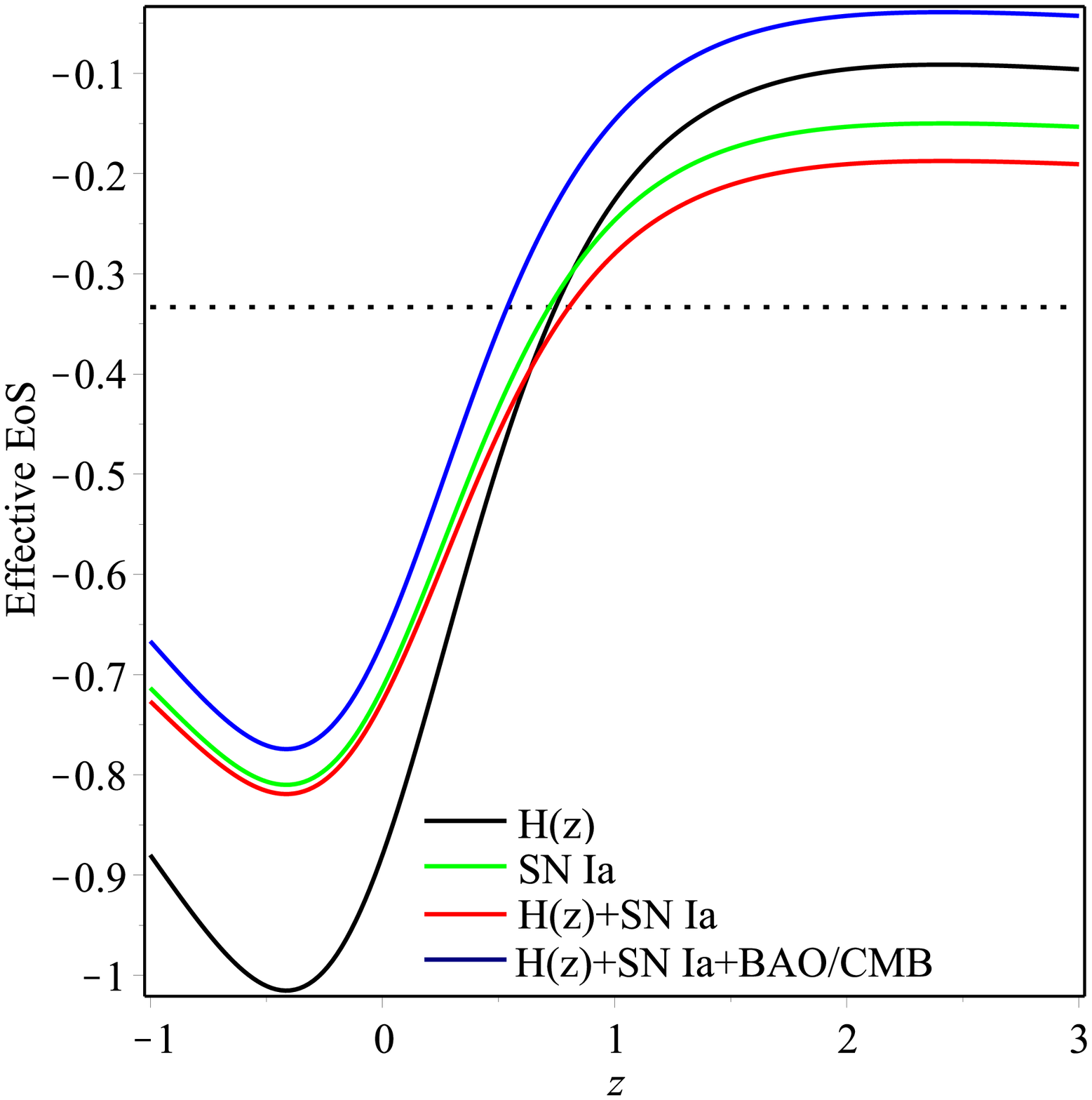}}\\
\subfigure[~$\Omega_{m}(z)$, $\Omega_{T}(z)$ evolutions]{\label{fig:Mod1-Om}\includegraphics[scale=.3]{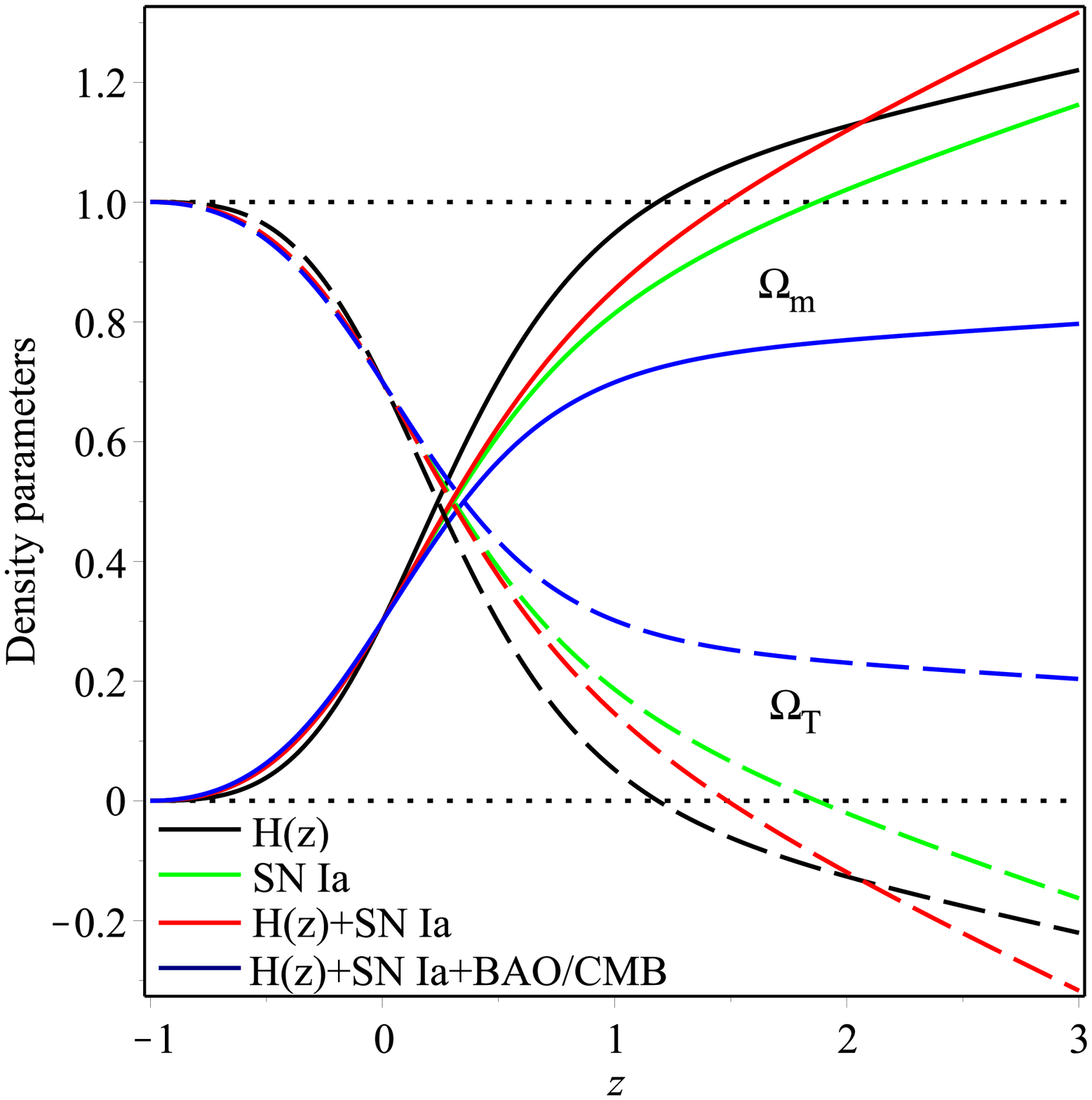}}
\subfigure[~$w_{T}$ evolution]{\label{fig:Mod1-wT}\includegraphics[scale=.3]{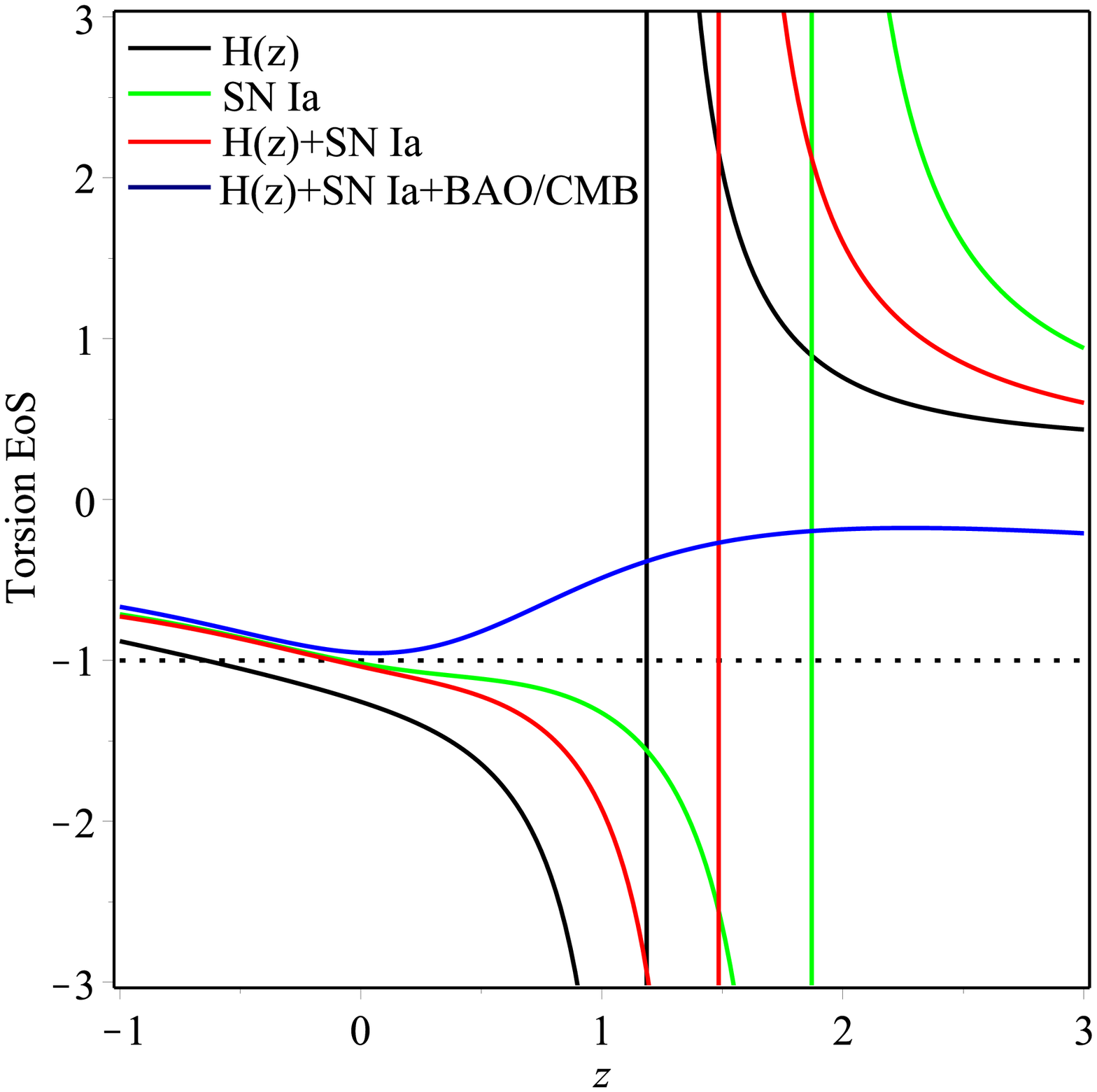}}
\caption[figtopcap]{\small{The best fit values of Model 1 parameters ($q_{0},~q_{1}$) are taken from \cite{Mamon:2015osa} according to the datasets combination used in that analysis, whereas ($-0.82,~0.98$) using $H(z)$ dataset, ($-0.57,~0.70$) using SN Ia dataset,  ($-0.59,~0.67$) using SN Ia + $H(z)$ datasets combination and ($-0.5,~0.78$) using SN Ia + $H(z)$ + BAO/CMB datasets combination, respectively;
\subref{fig:Mod1-fz} Evolution of $f(T(z))$ gravity (\ref{Mod1-f(z)}) normalized to $\Lambda$CDM (\ref{fLCDM}). For a viable theory, one expects it to oscillate about $\Lambda$CDM. As clear the theory is not in agreement with $\Lambda$CDM, therefore in practice one do not expect a viable thermal history;
\subref{fig:Mod1-weff} The effective (total) EoS does not match the standard cold dark matter (sCDM), $w_{eff}\to 0$, at redshifts $z\gtrsim 3$;
\subref{fig:Mod1-Om} For $H(z)$, SN Ia and SN Ia + $H(z)$ datasets, The matter density parameter crosses the unit boundary at redshifts $1 \lesssim z \lesssim 2$, while for SN Ia + $H(z)$ + BAO/CMB it occurs at $z\sim 10.3$;
\subref{fig:Mod1-wT} For $H(z)$, SN Ia and SN Ia + $H(z)$ datasets, the torsion (DE) EoS shows phase transition at redshifts $1 \lesssim z \lesssim 2$ as $w_T\to \pm \infty$, for SN Ia + $H(z)$ + BAO/CMB it diverges at $z\sim 10.3$. However, the theory shows better results when the CMB/BAO datasets are added, but it still cannot produce a thermal history compatible with the standard cosmology.}}
\label{Fig:Mod1}
\end{figure*}
At large redshift $z\gg 1$, the deceleration parameter $q(z)\to q_0+q_1$, which is suitable to study the radiation era. At late universe $0\leq z \ll 1$, the deceleration parameter reduces to the linear parametric form $q(z)=q_0+q_1 z$, so it is suitable to study the late accelerated expansion phase. Also, it can be shown that the deceleration parameter does not diverges as $z\to -1$, so it is suitable to study the fate of the universe. Since the above parametric form is finite for all redshift values $z\in \left[\right.-1, \infty\left.\right)$, it is valid to describe the entire cosmic history as mentioned in \cite{Mamon:2015osa}. Using the parametric form (\ref{Mod1-Xz}) and (\ref{Hubble-deceleration}), the Hubble-redshift relation can be written as
\begin{equation}\label{Mod1-H}
    H(z)=H_{0}(1+z)^{1+q_{0}}(1+z^2)^{\frac{q_{1}}{2}}.
\end{equation}
Additionally, by using the reconstruction equation (\ref{Reconstruction1}), we obtain the $f(z)$ form which controls the gravity sector
\begin{eqnarray}
\nonumber    f(z)&=&-6\Omega_{m,0}H_0^2(1+z)^{1+q_{0}}(1+z^2)^{\frac{q_{1}}{2}}\\
&&\times\int_0^z {\frac{1+q_{0}+q_{1}\frac{\bar{z}(1+\bar{z})}{1+\bar{z}^2}}{(1+\bar{z})^{q_{0}-1}(1+\bar{z}^{2})^{\frac{q_{1}}{2}}}} d\bar{z}.\label{Mod1-f(z)}
\end{eqnarray}
In Fig. \ref{Fig:Mod1}\subref{fig:Mod1-fz}, we plot the obtained $f(T)$ gravity verses the redshift according to different values of $q_0$ and $q_1$. The plots show that the theory has large deviations from the $\Lambda$CDM cosmology which is not favored in practice. This should be reflected on dynamical cosmological parameters like the matter density parameter.
\begin{figure*}[t!]
\centering
\subfigure[~$f(T)$ gravity]{\label{fig:Mod1A-fz}\includegraphics[scale=0.3]{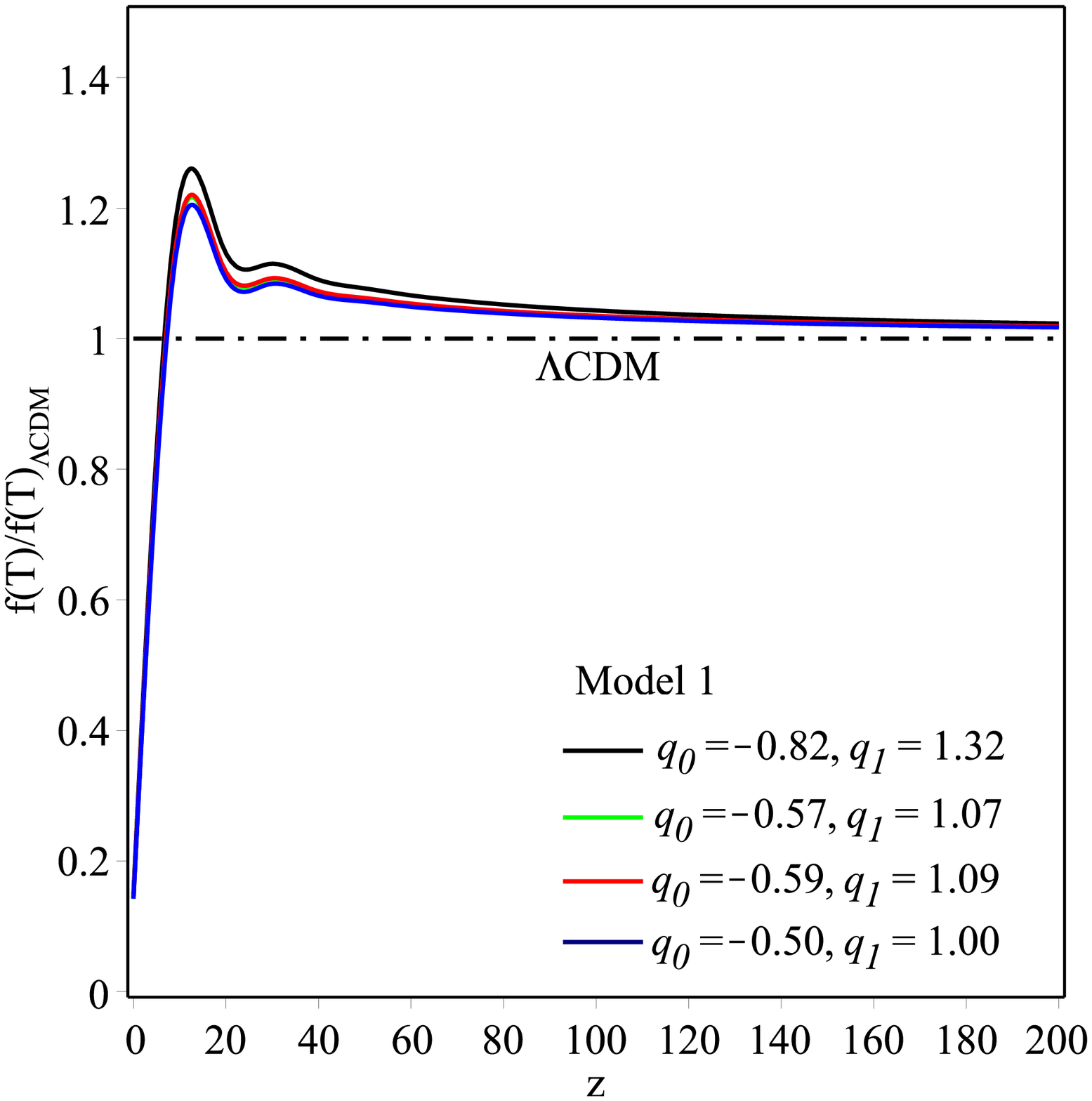}}
\subfigure[~$w_{eff}$ evolution]{\label{fig:Mod1A-weff}\includegraphics[scale=.3]{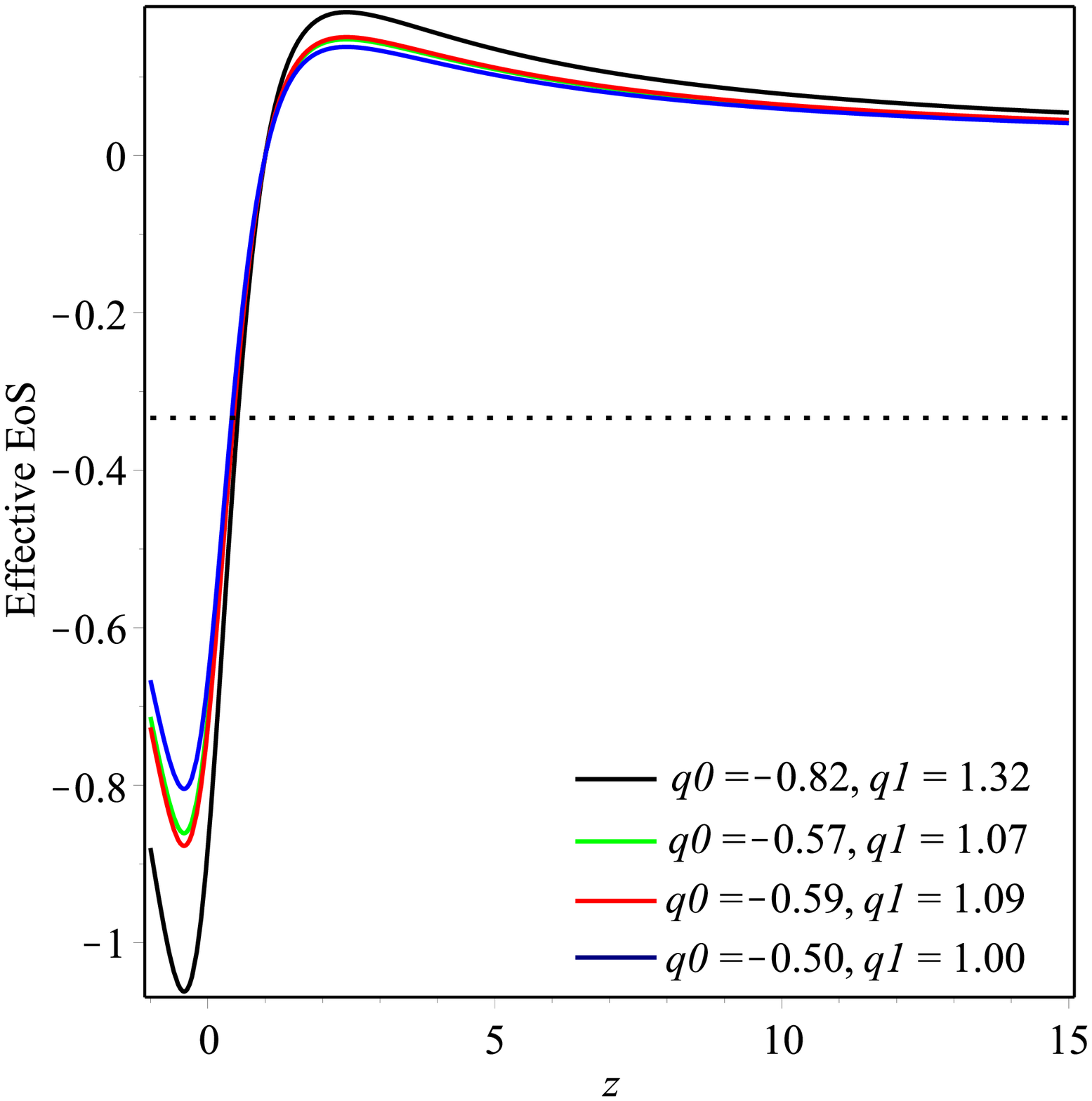}}\\
\subfigure[~$\Omega_{m}$, $\Omega_{T}$ evolution]{\label{fig:Mod1A-Om}\includegraphics[scale=.3]{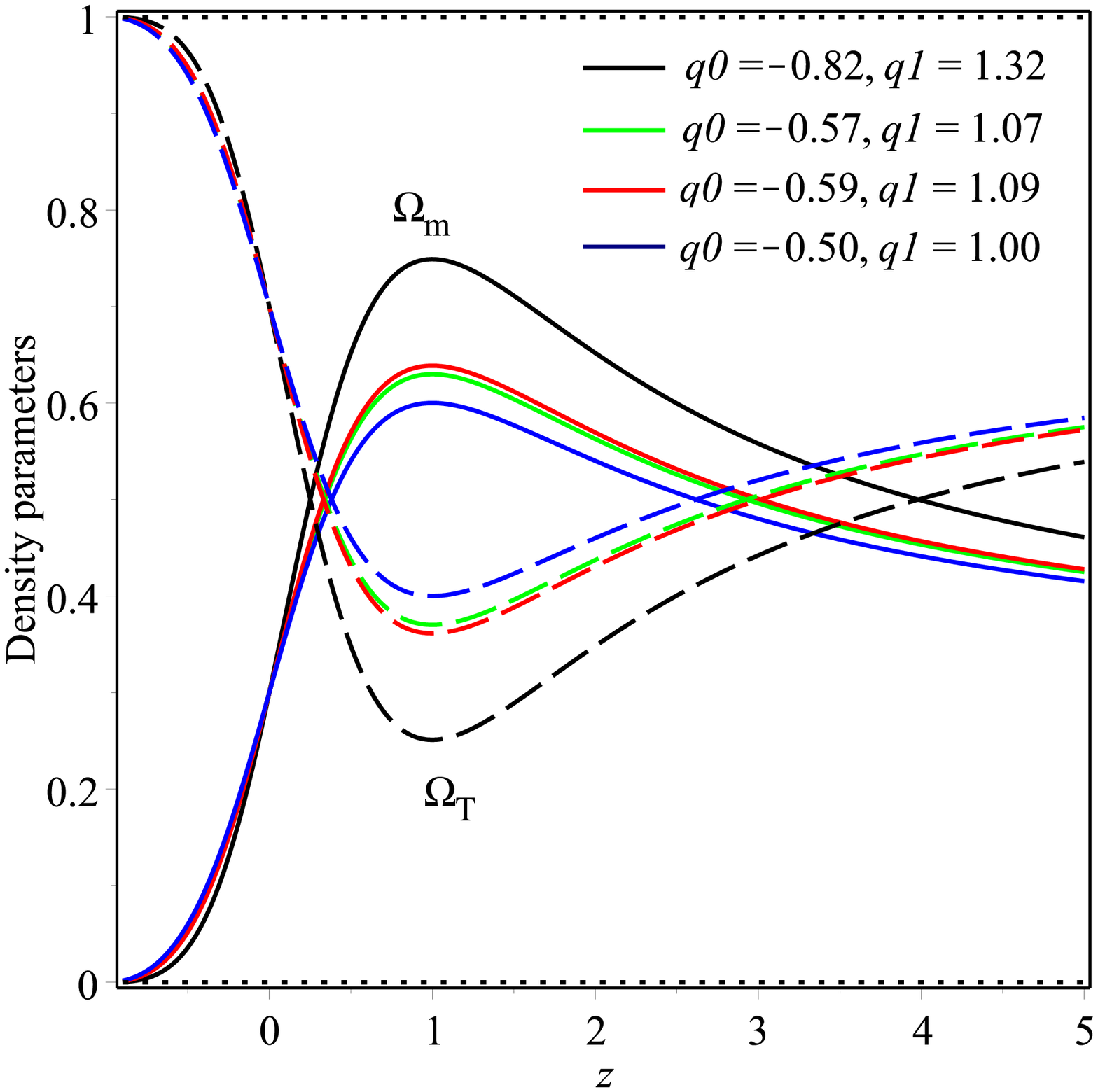}}
\subfigure[~$w_{T}$ evolution]{\label{fig:Mod1A-wT}\includegraphics[scale=.3]{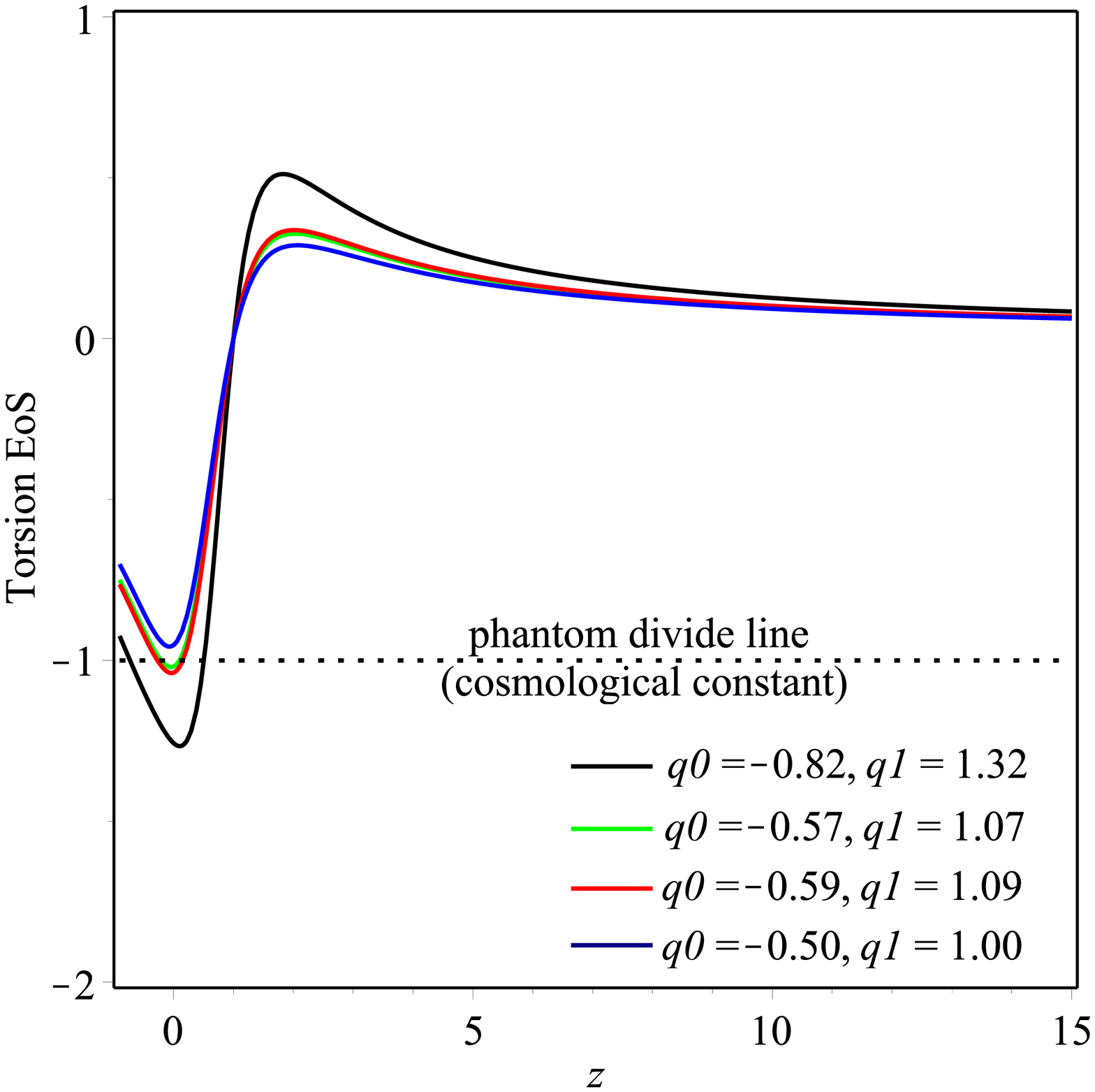}}
\caption[figtopcap]{\small{The best fit values of Model 1 parameters ($q_{0},~q_{1}$) are taken according to the constraint (\ref{q1-min}). The model parameter $q_0$ is kept fixed to its value as measured in \cite{Mamon:2015osa}, since it is compatible with the present values of other cosmological parameters. However, the model parameter $q_1$ is recalculated to fulfill (\ref{q1-min}) as given on the plots;
\subref{fig:Mod1A-fz} The $f(T)$ gravity matches $\Lambda$CDM at large redshifts on the contrary to the corresponding plots in Fig. \ref{Fig:Mod1}\subref{fig:Mod1-fz}.
\subref{fig:Mod1A-weff} The effective (total) EoS shows that the universe can effectively produce sCDM dominant era at large redshifts as $w_{eff}\to 0$;
\subref{fig:Mod1A-Om} The matter density parameter does not cross the uint boundary line anymore and consequently the torsional density parameter does not have negative values;
\subref{fig:Mod1A-wT} The torsion (DE) EoS has finite values at all redshifts.}}
\label{Fig:Mod1A}
\end{figure*}

We next evaluate the total EoS parameter according to the parametrization (\ref{Mod1-Xz}). Then Eq. (\ref{deceleration}) reads
\begin{equation}\label{Mod1-weff}
    w_{eff}(z)=-1+\frac{2}{3}\frac{(1+q_{0})+q_{1}z+(1+q_{0}+q_{1})z^2}{1+z^2}.
\end{equation}
According to the values of the parameters $q_0$ and $q_1$ given in \cite{Mamon:2015osa}, the transition redshift $z_{tr}$ can be determined by setting $w_{eff}(z_{tr})=-1/3$. This gives $z_{tr}\backsimeq 0.75$, $0.72$, $0.8$ and $0.54$ according to the datasets $H(z)$, SNIa, $H(z)$+SNIa and $H(z)$+SNIa+BAO/CMB, respectively. The evolution of $w_{eff}(z)$ is given as in Fig. \ref{Fig:Mod1}\subref{fig:Mod1-weff}. Although the plots show transition redshift in agreement with observations, they are clearly incompatible with the sCDM behavior (i.e $w_{eff}(z)=0$) at large redshift. This confirms the inefficiency of the model at the earlier phases (large redshifts).
\begin{table*}[t!]
\caption{\label{Table1}%
The main results of model 1 according to the values of ($q_0,q_1$) parameters as given in Ref. \cite{Mamon:2015osa} and by using the matter density parameter constraint (\ref{q1-min}). In the later, we keep the strict measured values $q_0$ as they are, while choosing the corresponding values of $q_1$ to fulfill the constraint.}
\begin{ruledtabular}
\begin{tabular*}{\textwidth}{lccccccc}
\multirow{2}{*}{\textbf{Dataset}}                        & \multirow{2}{*}{$q_0$} && \multirow{2}{*}{$q_1$} & $f(T)/\Lambda$CDM  & $\Omega_m(z)\leq 1$          & \textbf{Torsion} & \multirow{2}{*}{\textbf{Viability}} \\
                                        &       &&       & \textbf{compatibility} & \textbf{constraint} & \textbf{EoS}, $\omega_T$           &  \\
\hline
H(z)                                    &       $-0.82$      &&      $0.98$       & not                          & violated ($z\sim 1.24$)                             & diverges\footnote{\label{footnote:1a}Note that the torsion EoS diverges when the matter density parameter exceeds the unit boundary line (equivalently, when the torsion density parameter becomes negative), see Figs. \ref{Fig:Mod1}\subref{fig:Mod1-Om} and \ref{Fig:Mod1}\subref{fig:Mod1-wT}.} ($z\sim 1.24$)         & not                 \\
SNIa                                    &     $-0.57$        &&       $0.70$      & not                          & violated ($z\sim1.84$)                             & diverges$^\textrm{\ref{footnote:1a}}$ ($z\sim1.84$)         & not                 \\
H(z)+SNIa                               &       $-0.59$      &&      $0.67$       & not                         & violated ($z\sim1.45$)                             & diverges$^\textrm{\ref{footnote:1a}}$ ($z\sim1.45$)        & not                 \\
H(z)+SNIa+BAO/CMB                       &        $-0.50$     &&       $0.78$      & not                          & violated ($z\sim 10.28$)                            & diverges$^\textrm{\ref{footnote:1a}}$ ($z\sim 10.28$)    & not                 \\
\hline
\textbf{Using constraint (\ref{q1-min})}    & \textbf{}   && \textbf{}   & \textbf{}                   & \textbf{}                      & \textbf{}        & \textbf{}          \\
\hline
H(z)              &      $-0.82$       &&     $1.32$        & semi                        & fulfilled                            & does not diverge     & not\footnote{\label{footnote:1b}Although the enhanced parameters-using the matter density parameter constraint (\ref{q1-min})- give $f(T)$ gravity with better compatibility with $\Lambda$CDM and smooth $\omega_T$, it cannot produce viable patterns of the matter density parameter as seen in Fig. \ref{Fig:Mod1A}\subref{fig:Mod1A-Om}.}                \\
SNIa              &       $-0.57$       &&     $1.07$        & semi                        & fulfilled                            & does not diverge     & not$^\textrm{\ref{footnote:1b}}$                \\
H(z)+SNIa         &        $-0.59$      &&     $1.09$        & semi                        & fulfilled                            & does not diverge     & not$^\textrm{\ref{footnote:1b}}$                \\
H(z)+SNIa+BAO/CMB &       $-0.50$      &&       $1.00$      & semi                        & fulfilled                            & does not diverge     & not$^\textrm{\ref{footnote:1b}}$
\end{tabular*}
\end{ruledtabular}
\end{table*}

Using the parametrization (\ref{Mod1-Xz}), the matter density parameter (\ref{matter-density-parameter}) reads
\begin{equation}\label{Mod1-Omega_m}
    \Omega_{m}(z)=\Omega_{m,0}(1+z)^{1-2q_{0}}(1+z^2)^{-q_{1}}.
\end{equation}
In Fig. \ref{Fig:Mod1}\subref{fig:Mod1-Om}, we plot the evolution of the matter and the torsion density parameters according to the calculated values of the model parameters $q_0$ and $q_1$ as given in Ref. \cite{Mamon:2015osa}. As shown by the plots, the matter density parameter crosses the unit boundary line at redshift $1\lesssim z \lesssim2$ for the datasets $H(z)$, SNIa and $H(z)$+SNIa, while for the combination $H(z)$+SNIa+BAO/CMB, the matter density parameter $\Omega_{m}(z)$ crosses the unity at larger redshift $z\gtrsim 10$. As expected from the analysis of the obtained $f(T)$ gravity, namely Eq. (\ref{Mod1-f(z)}), the parametrization (\ref{Mod1-Xz}) does not produce a sCDM compatible with the thermal history.

Also we evaluate the torsional EoS parameter associated to the parametric form (\ref{Mod1-Xz}),
\begin{equation}\label{Mod1-wDE}
    w_{T}(z)=\frac{2(1+z)^{2q_{0}}\left[(q_{0}-\frac{1}{2})+\frac{q_{1}z(1+z)}{1+z^2}\right]}
    {3\left[(1+z)^{2q_{0}}-\frac{\Omega_{0,m}(1+z)}{(1+z^2)^{q_{1}}}\right]}.
\end{equation}
According to the values of the parameters $q_0$ and $q_1$ given in \cite{Mamon:2015osa}, we plot the evolution of $w_T(z)$ in Fig. \ref{Fig:Mod1}\subref{fig:Mod1-wT}. The plots show that the torsional EoS parameter evolves in phantomlike regime at low redshifts. We note that a possible phase transition occurs at redshifts $z\sim 1.19$, $z\sim 1.87$, $z\sim 1.47$ and $z\sim 10.29$ for the datasets $H(z)$, SNIa, $H(z)$+SNIa and $H(z)$+SNIa+BAO/CMB as $w_{T}\to \pm \infty$. Also the model forecasts crossing the phantom divide line at future to quintessencelike regime. Remarkably, the phase transitions of torsion gravity is associated to crossing the matter density parameter, $\Omega_{m}(z)$, the unit boundary line (or when $\Omega_T$ crosses to negative region) as clear in Figs. \ref{Fig:Mod1}\subref{fig:Mod1-Om} and \ref{Fig:Mod1}\subref{fig:Mod1-wT}.

In the following part of this subsection, we show how the model parameters can be constrained aiming to produce viable cosmic evolution. This means that if the predicted values of the model parameters agree with their measured ones, the assumed parametrization could be a good approximation to describe the cosmic history. As a matter of fact, we need the matter density parameter to reach a maximal value $\Omega_{m}(z)=1$ asymptotically, i.e as $z\to \infty$. This condition is useful to add a further constraint on the free parameters $q_0$ and $q_1$. In more detail, we write the asymptotic expansion of the matter density parameter (\ref{Mod1-Omega_m}) up to the second order of the redshift
$$\widetilde{\Omega}_{m}(z)\thickapprox \Omega_{m,0} \left(\frac{1}{z}\right)^{2(q_{0}+q_{1})}(1+z-2q_0)+O(1/z^2).$$
For viable models, we need $\widetilde{\Omega}_{m}(z)\leq 1$, otherwise the torsion density parameter should drop to negative values. This constrains $q_{1}$ to a minimum value
\begin{equation}\label{q1-min}
    q_{1}\geq \lim_{z\to \infty}\frac{\ln \left[z^{-2q_{0}}(1+z-2q_{0})\Omega_{0,m}\right]}{\ln z^{2}} = \frac{1}{2}-q_{0}.
\end{equation}
The above inequality sets a constraint on the choice of the model parameters, that is
$$q_{0}+q_{1}\geq 0.5\,.$$
If $q_{0}+q_{1}<0.5$, the matter density parameter would exceed the unity at some redshift at past. The closer $q_{0}+q_{1} \to 0.5^{-}$, the earlier $\Omega_{m}(z)$ crossing to the unit boundary, i.e. at larger $z$. This can easily seen from Fig. \ref{Fig:Mod1}\subref{fig:Mod1-Om}, whereas the sums of the pairs $q_0+q_1$ are $0.16$, $0.13$ and $0.08$ for the datasets $H(z)$, SNIa and $H(z)$+SNIa, respectively. For the combination $H(z)$+SNIa+BAO/CMB, the sum $q_0+q_1=0.28$ which is closer to the critical value $0.5$, and therefore the matter density parameter crosses the unit boundary at larger redshift $z\sim 10$. However, we find that these patterns are not compatible with the $\Lambda$CDM behavior as mentioned before.

In order to enhance the model predictions, we fix the $q_0$-value as measured in \cite{Mamon:2015osa}, since its value is compatible with the present values of other cosmological parameters. However, we use the matter domination condition (\ref{q1-min}) to recalculate the corresponding $q_1$ value for each dataset. In this case, we enhance the evolution $f(T)$ gravity as shown in Fig. \ref{Fig:Mod1A}\subref{fig:Mod1A-fz}. By comparison with Fig. \ref{Fig:Mod1}\subref{fig:Mod1-fz}, we find that the plots match the $\Lambda$CDM model at large redshifts. On the other hand, the universe effectively can describe sCDM matter domination era as $w_{eff}\to 0$ at large redshifts, see Fig. \ref{Fig:Mod1A}\subref{fig:Mod1A-weff}. Indeed, the matter density parameter does not cross the unit boundary line, but the behavior still inconsistent with the matter domination era. This can be shown clearly in Fig. \ref{Fig:Mod1A}\subref{fig:Mod1A-Om}. Finally, we note that the torsional EoS in the enhanced version does not indicate phase transitions anymore, whereas $w_T$ becomes finite at all redshifts, see Fig. \ref{Fig:Mod1A}\subref{fig:Mod1A-wT}. In Table \ref{Table1}, we summarize the results of model 1 according to the values of the model parameters $q_0$ and $q_1$ as given in Ref.  \cite{Mamon:2015osa} and by applying the matter density constraint (\ref{q1-min}).

In conclusion, the model parameters in the enhanced version neither match the best fit values as measured by different datasets combinations nor produce a viable behavior of the matter density parameter. Therefore, we note that although the parametric form (\ref{Mod1-Xz}) of the deceleration parameter does not diverge in the full redshift range $z\in \left[\right.-1, \infty\left.\right)$ it cannot be considered as a viable model to produce the whole cosmic history.

\subsection{Model 2}\label{Sec4.3}
In this subsection, we examine another $q(z)$-parametrization \cite{Mamon:2016dlv}
\begin{equation}\label{Mod2-Xz}
    X(z)=\frac{\ln{(N+z)}}{(1+z)}-\ln{N};\quad N>1.
\end{equation}
\begin{figure*}[t!]
\centering
\subfigure[~$f(T)$ gravity evolution]{\label{fig:Mod2-fz}\includegraphics[scale=0.3]{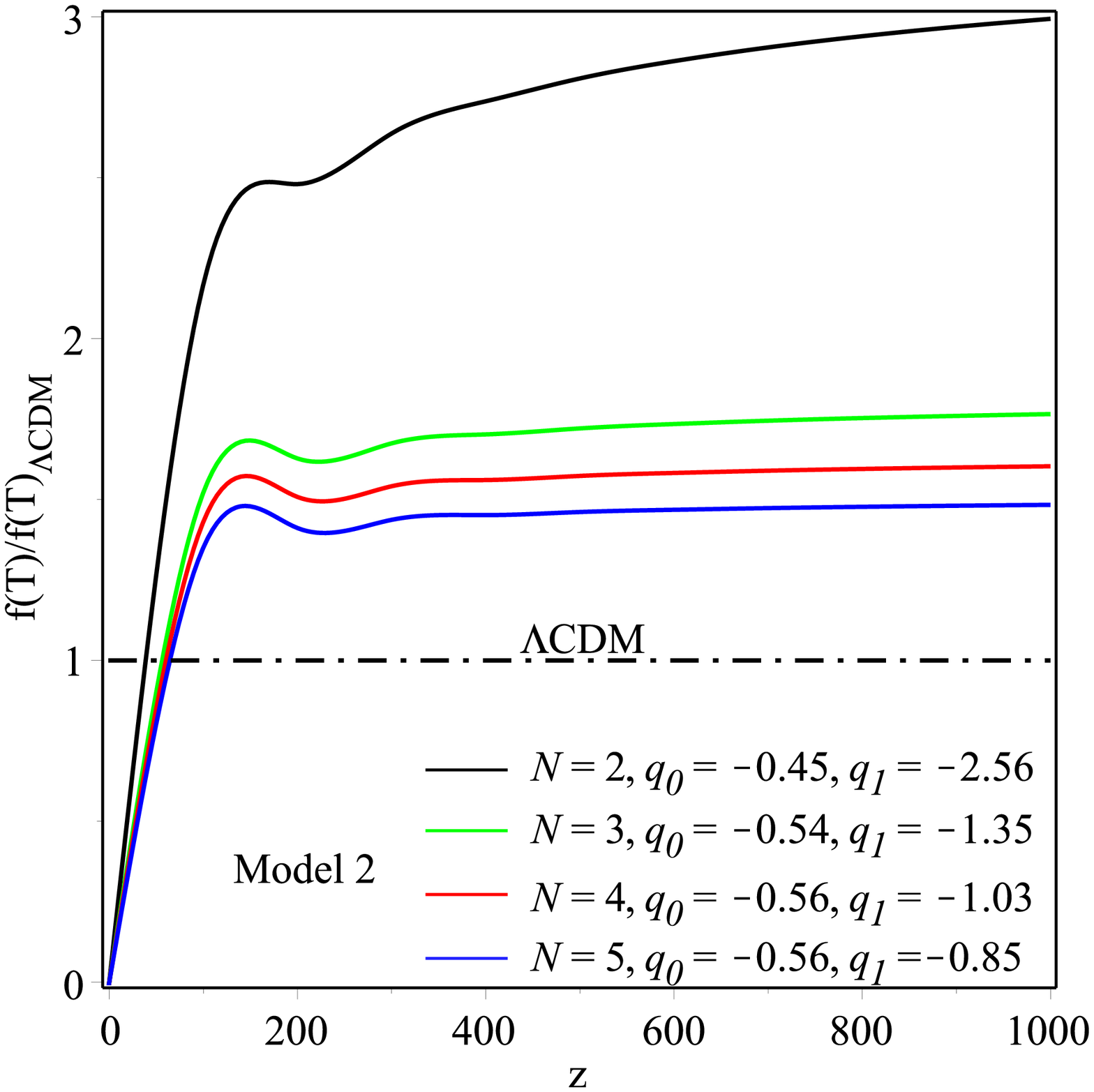}}
\subfigure[~$w_{eff}$ evolution]{\label{fig:Mod2-weff}\includegraphics[scale=.3]{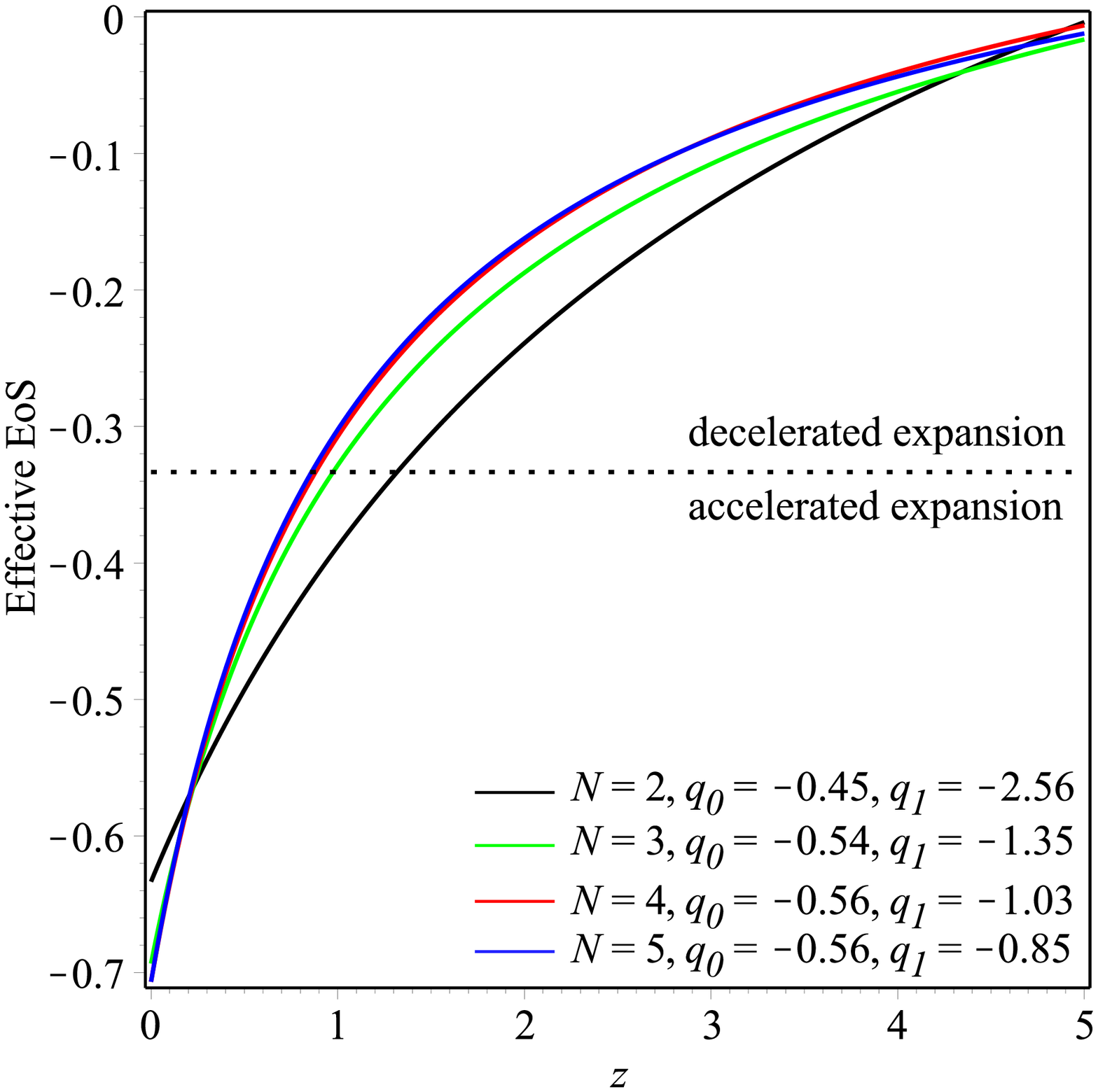}}\\
\subfigure[~$\Omega_{m}(z)$, $\Omega_{T}(z)$ evolutions]{\label{fig:Mod2-Om}\includegraphics[scale=.3]{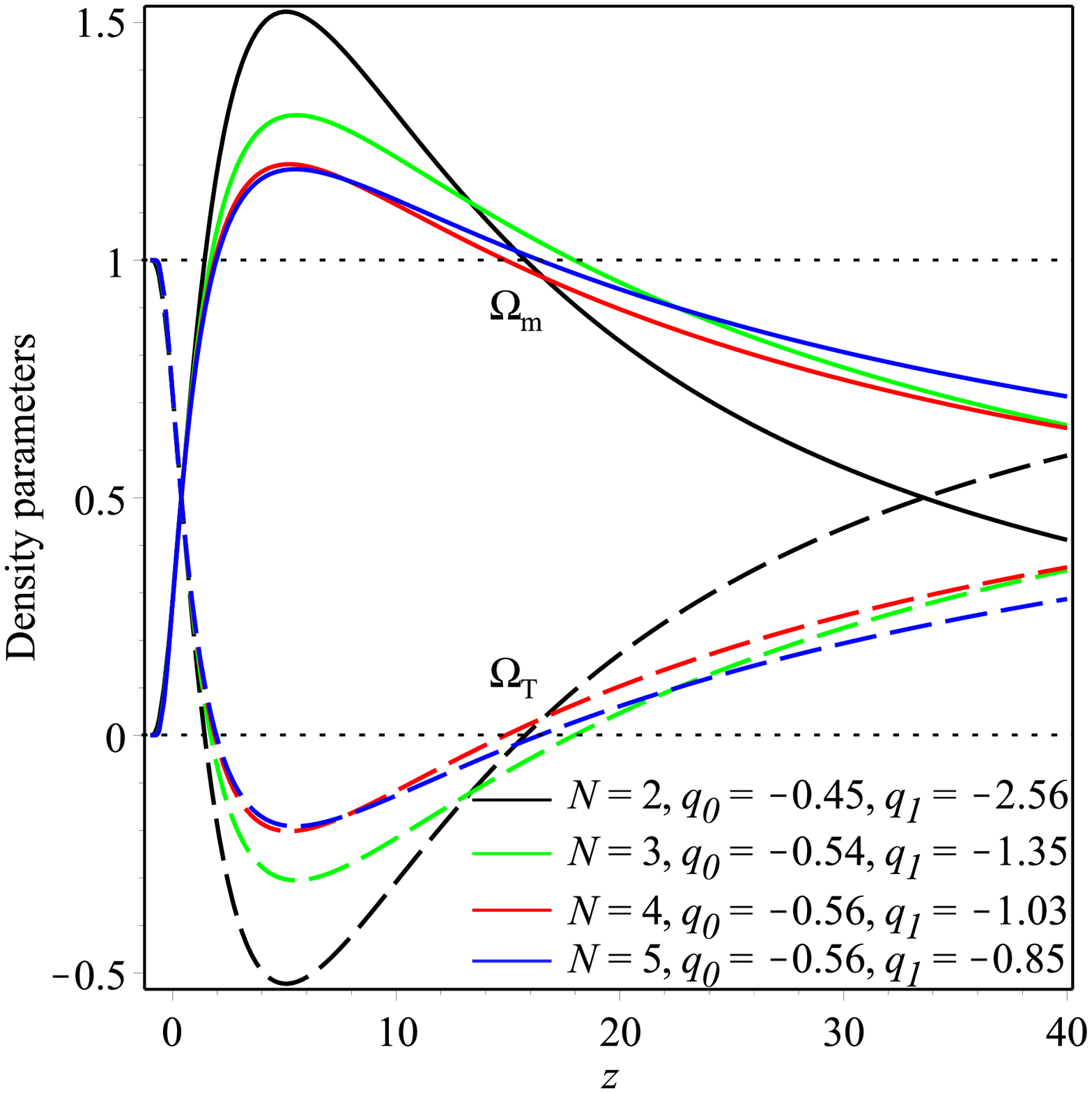}}
\subfigure[~$w_{T}$ evolution]{\label{fig:Mod2-wT}\includegraphics[scale=.3]{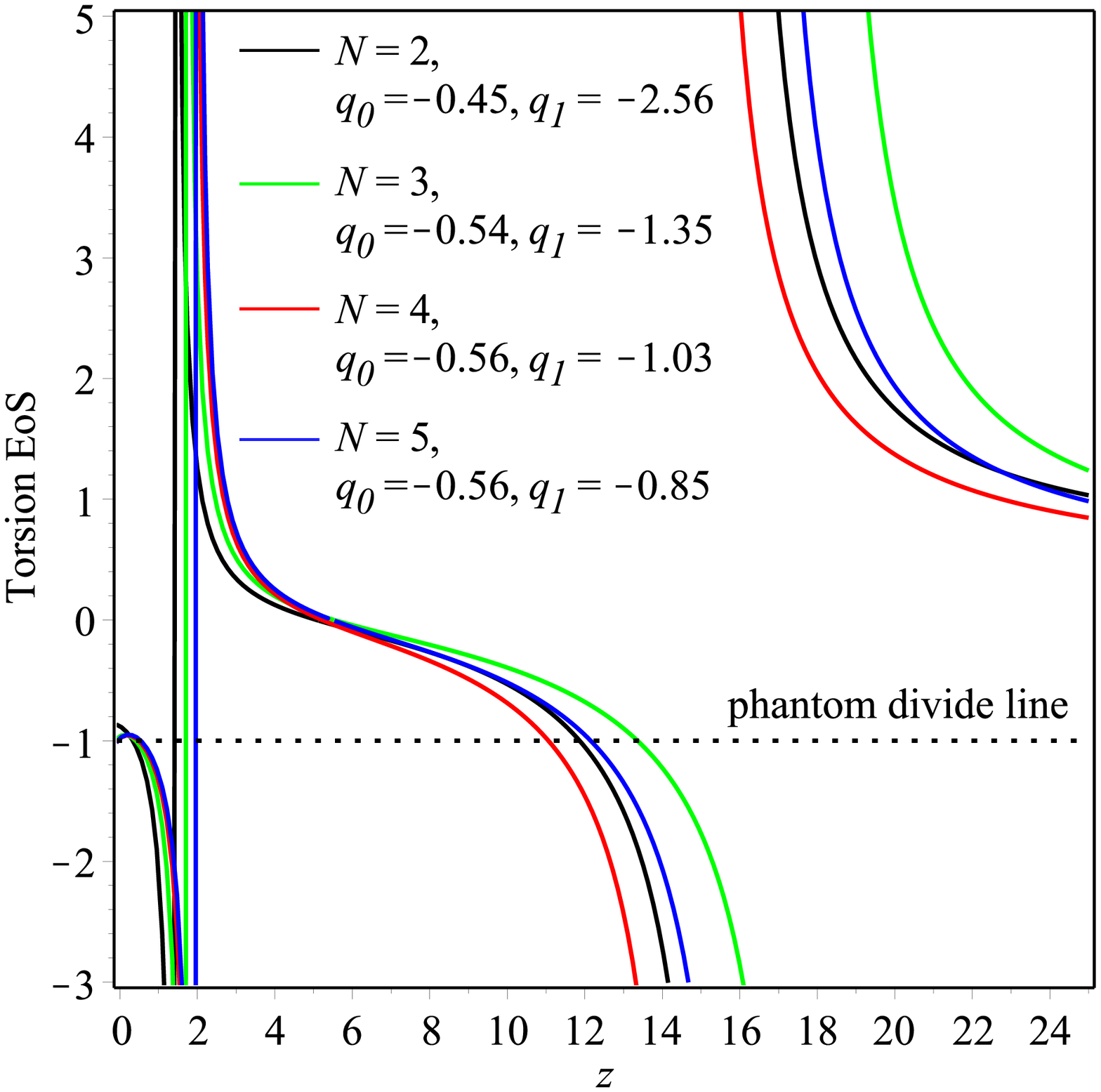}}
\caption[figtopcap]{\small{The best fit values of Model 2 parameters ($q_{0},~q_{1}$) are taken from \cite{Mamon:2016dlv} according to the datasets combination used in that analysis, whereas ($N = 2$, $q_0 = -0.45$, $q_1 = -2.56$), ($N = 3$, $q_0 = -0.54$, $q_1 = -1.35$), ($N = 4$, $q_0 = -0.56$, $q_1 = -1.03$) and ($N = 5$, $q_0 = -0.56$, $q_1 = -0.85$);
\subref{fig:Mod2-fz} Evolution of $f(T(z))$ gravity (\ref{Mod1-f(z)}) normalized to $\Lambda$CDM (\ref{fLCDM}), as clear the theory is not in agreement with $\Lambda$CDM so in practice one do not expect a viable thermal history;
\subref{fig:Mod2-weff} The effective (total) EoS does not match the sCDM, $w_{eff}\to 0$, at redshifts $z\gtrsim 3$;
\subref{fig:Mod2-Om} The matter density parameter crosses the unit boundary at redshifts $1 \lesssim z \lesssim 2$;
\subref{fig:Mod2-wT} The torsion (DE) EoS show phase transition at redshifts $1 \lesssim z \lesssim 2$ as $w_T\to \pm \infty$. However, the theory shows better results when the CMB/BAO datasets are added, but it still cannot produce a thermal history compatible with the standard cosmology.}}
\label{Fig:Mod2}
\end{figure*}
Using the above parametrization, the deceleration parameter (\ref{general-param-q}) has been confronted with observational datasets, in particular JLA SNIa and BAO/CMB, whereas the best fit values of the free parameters $q_0$ and $q_1$ have been calculated up to 1$\sigma$ \cite{Mamon:2016dlv}. For different choices of the parameter $N$, it has been shown that ($N = 2$, $q_0 = -0.45$, $q_1 = -2.56$), ($N = 3$, $q_0 = -0.54$, $q_1 = -1.35$), ($N = 4$, $q_0 = -0.56$, $q_1 = -1.03$) and ($N = 5$, $q_0 = -0.56$, $q_1 = -0.85$). In addition, it has been shown that, at present time $z=0$, the above parametrization leads the deceleration parameter to have a value $q=q_0$, while at large redshift one can obtain the matter dominant era, $q=\frac{1}{2}$, by using the constraint $q_1=\frac{2q_0-1}{2\ln{N}}$. Using the parametric form (\ref{Mod2-Xz}) and (\ref{Hubble-deceleration}), the Hubble-redshift relation can be written as
\begin{figure*}[t!]
\centering
\subfigure[~$f(T)$ gravity]{\label{fig:Mod2A-fz}\includegraphics[scale=0.3]{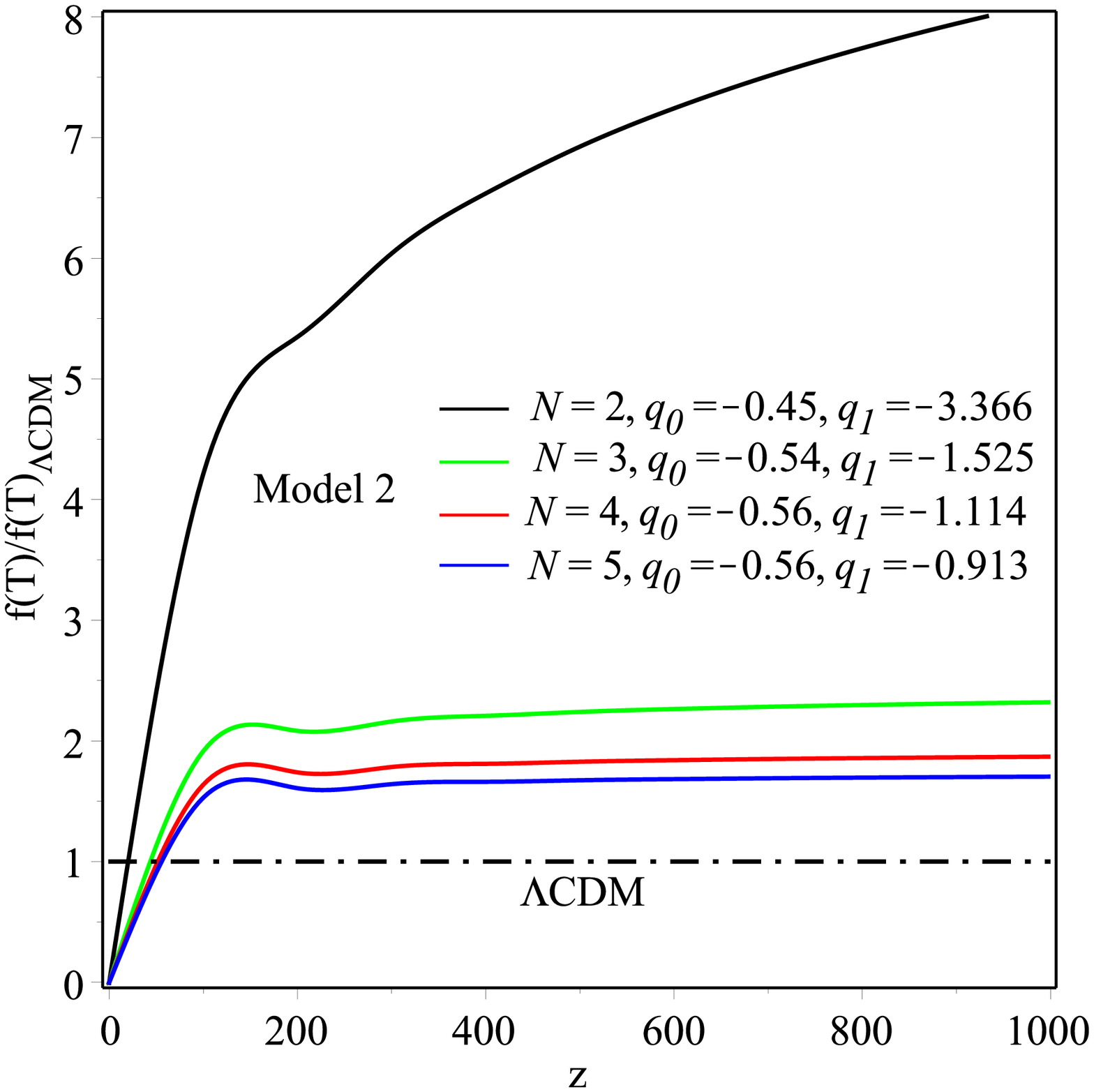}}
\subfigure[~$w_{eff}$ evolution]{\label{fig:Mod2A-weff}\includegraphics[scale=.3]{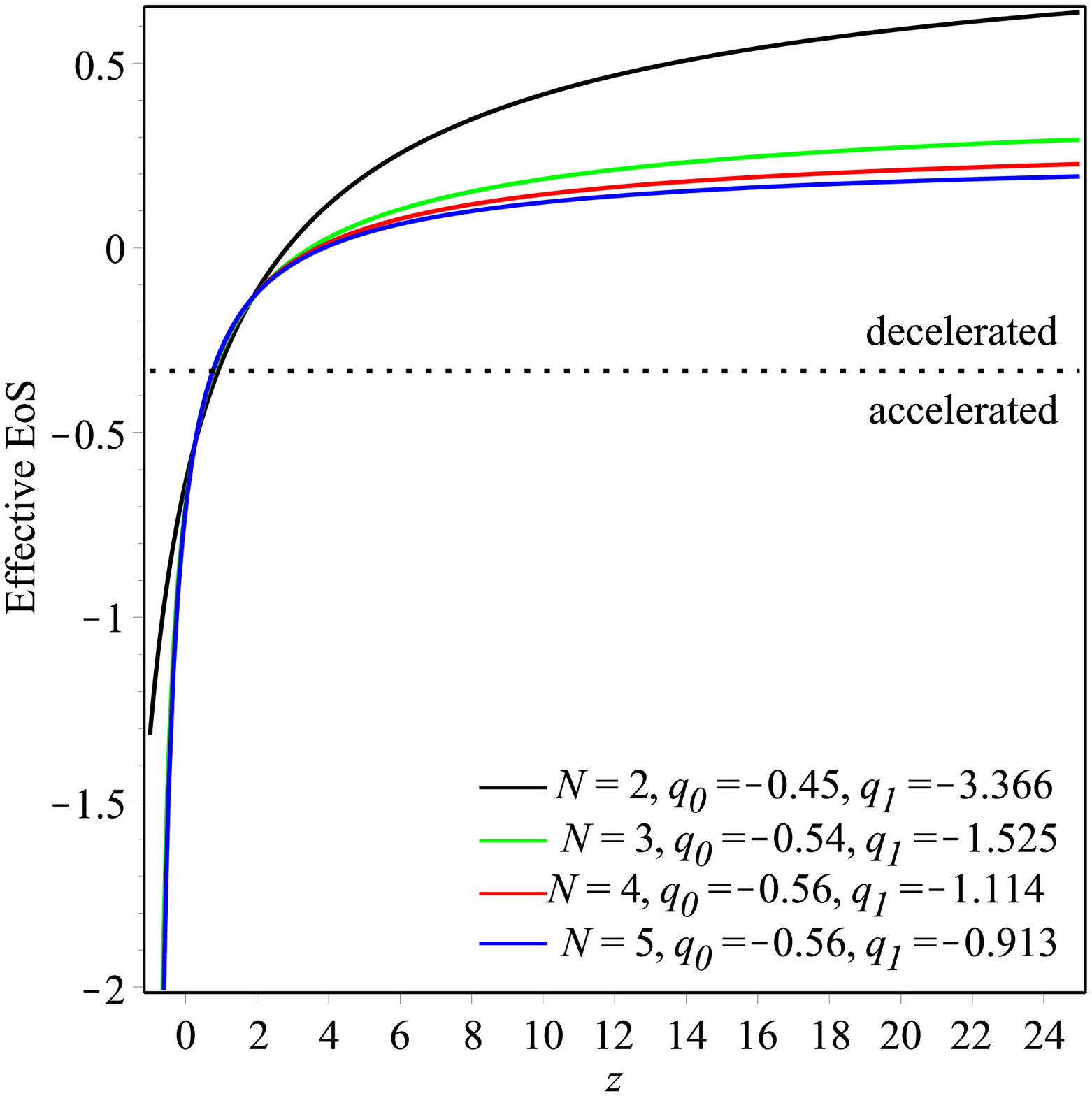}}\\
\subfigure[~$\Omega_{m}$, $\Omega_{T}$ evolution]{\label{fig:Mod2A-Om}\includegraphics[scale=.3]{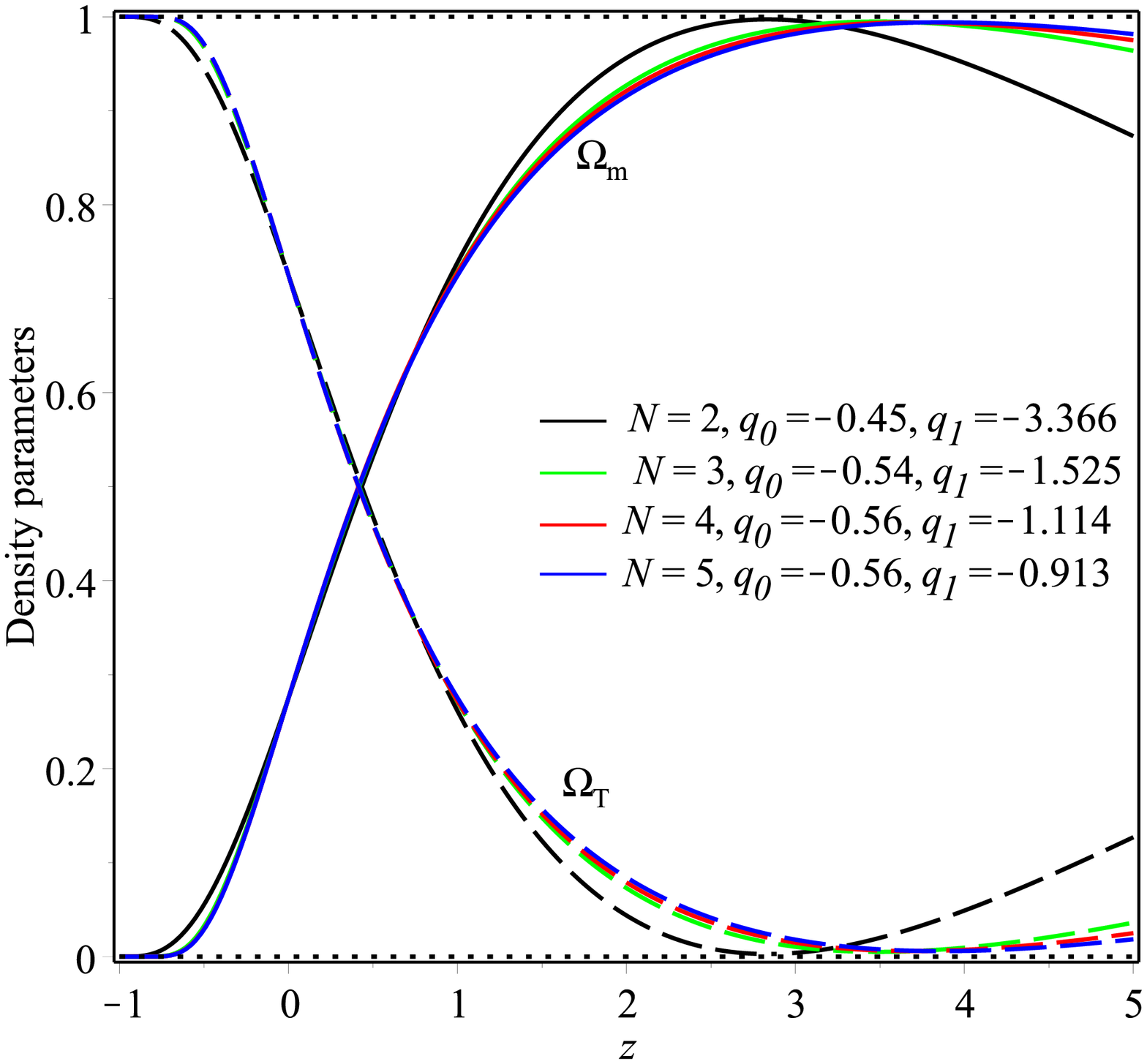}}
\subfigure[~$w_{T}$ evolution]{\label{fig:Mod2A-wT}\includegraphics[scale=.3]{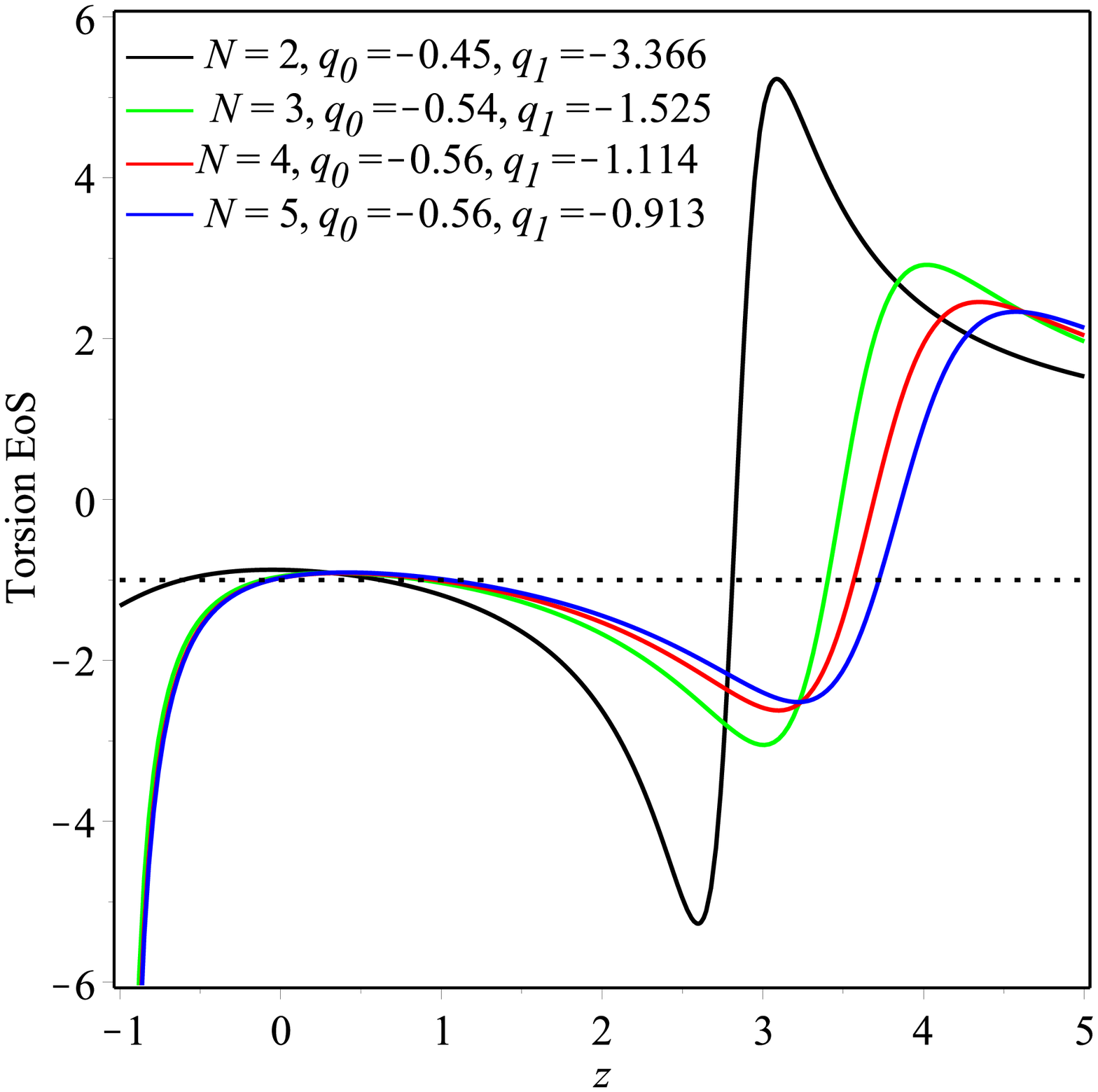}}
\caption[figtopcap]{\small{The best fit values of Model 2 parameters ($q_{0},~q_{1}$) are taken according to the constraint (\ref{q2-min}). The model parameter $q_0$ is kept fixed to it value as measured in \cite{Mamon:2016dlv}, since it is compatible with the present values of other cosmological parameters. However, the model parameter $q_1$ is recalculated to fulfill (\ref{q2-min}) as given on the plots;
\subref{fig:Mod2A-fz} The $f(T)$ gravity matches $\Lambda$CDM at large redshifts on the contrary to the corresponding plots in Fig. \ref{Fig:Mod2}\subref{fig:Mod2-fz}.
\subref{fig:Mod2A-weff} The effective (total) EoS shows that the universe can effectively produce sCDM dominant era at large redshifts as $w_{eff}\to 0$;
\subref{fig:Mod2A-Om} The matter density parameter does not cross the unit boundary line anymore and consequently the torsional density parameter does not have negative values;
\subref{fig:Mod2A-wT} The torsion (DE) EoS has finite values at all redshifts.}}
\label{Fig:Mod2A}
\end{figure*}
\begin{equation}\label{Mod2-H}
    H(z)=H_{0}N^r(1+z)^s(N+z)^{-\frac{(N+z)r}{(1+z)N}},
\end{equation}
where $r:={\frac{q_1 N}{N-1}}$ and $s:=1+q_0+\frac{r}{N}-q_1\ln{N}$. Additionally, by using the reconstruction equation (\ref{Reconstruction1}), we obtain the $f(z)$ form which controls the gravity sector
\begin{eqnarray}
    f(z)&=&-6\Omega_{m,0}H_0^2\frac{(1+z)^s}{(N+z)^{\frac{(N+z)r}{(1+z)N}}} \label{Mod2-f(z)}\\
&&\times    \int_0^z\frac{(1+\bar{z})^{2-s}\left(1+q_0+q_1 \frac{\ln{(N+\bar{z})}}{(1+\bar{z})}-\ln{N}\right)}{(N+\bar{z})^{-\frac{(N+\bar{z})r}{(1+\bar{z})N}}}~d\bar{z}.\nonumber
\end{eqnarray}
In Fig. \ref{Fig:Mod2}\subref{fig:Mod2-fz}, we plot the obtained $f(T)$ gravity verses the redshift according to different values of $q_0$ and $q_1$ as measured in \cite{Mamon:2016dlv}. Similar to Model 1, the plots show that the theory has large deviations from the $\Lambda$CDM cosmology which is not favored in practice. This should be reflected on dynamical cosmological parameters like the matter density parameter.

We next evaluate the total EoS parameter according to the parametrization (\ref{Mod2-Xz}). Then Eq. (\ref{deceleration}) reads
\begin{equation}\label{Mod2-weff}
    w_{eff}(z)=-\frac{1}{3}+\frac{2}{3}\left[q_0+q_1\left(\frac{\ln{(N+z)}}{(1+z)}-\ln{N}\right)\right].
\end{equation}
According to the values of the parameters $q_0$ and $q_1$ given in \cite{Mamon:2016dlv}, the transition redshift $z_{tr}$ can be determined by setting $w_{eff}(z_{tr})=-1/3$. This gives $z_{tr}\backsimeq 1.32$, $0.98$, $0.88$ and $0.86$ according to the datasets SNIa+BAO/CMB with different choices of $N=2\cdots 5$, respectively. The evolution of $w_{eff}(z)$ is given as in Fig. \ref{Fig:Mod2}\subref{fig:Mod2-weff}. Although the plots show transition redshift in agreement with observations, they are clearly incompatible with the sCDM behavior (i.e $w_{eff}(z)=0$) at large redshift. This confirms the inefficiency of the model at the earlier phases (large redshifts).

Using the parametrization (\ref{Mod2-Xz}), the matter density parameter (\ref{matter-density-parameter}) reads
\begin{equation}\label{Mod2-Omega_m}
    \Omega_{m}(z)=\frac{\Omega_{m,0}}{N^{2r}} (1+z)^{3-2s} (N+z)^{2\frac{(N+z)r}{(1+z)N}}.
\end{equation}
In Fig. \ref{Fig:Mod2}\subref{fig:Mod2-Om}, we plot the evolution of the matter and the torsion density parameters according to the calculated values of the model parameters $q_0$ and $q_1$ as given in Ref. \cite{Mamon:2016dlv}. As shown by the plots, the matter density parameter $0<\Omega_{m}(z)<1$ at large redshifts, then it peaks up at low redshifts crossing the unit boundary line at redshifts $z\sim 15.78$, $17.96$, $14.89$ and $16.41$ for the values $N=2\cdots 5$, respectively. However, the matter density drops down once again crossing the unit boundary line at redshifts $z\sim 1.43$, $1.70$, $1.88$ and $1.95$ for the values $N=2\cdots 5$, respectively. In agreement with the analysis of the obtained $f(T)$ gravity, namely Eq. (\ref{Mod2-f(z)}), the parametrization (\ref{Mod2-Xz}) does not produce a matter dominant era compatible with the thermal history.

Also, we evaluate the torsional EoS parameter associated to the parametric form (\ref{Mod2-Xz}),
\begin{equation}\label{Mod2-wDE}
    w_{T}(z)=\frac{-1+2q_0+2q_1\left[\frac{\ln{N+z}}{1+z}-\ln{N}\right]}{3-3\Omega_{m,0}N^r(1+z)^{3-2s}(N+z)^{\frac{2r(N+z)}{N(1+z)}}}.
\end{equation}
According to the values of the parameters $q_0$ and $q_1$ given in \cite{Mamon:2016dlv}, we plot the evolution of $w_T(z)$ in Fig. \ref{Fig:Mod2}\subref{fig:Mod2-wT}. The plots show that the torsional EoS parameter evolves in phantomlike regime at low redshifts. We note that an early phase transition occurs at redshifts $z\sim 15.78$, $17.96$, $14.89$ and $16.41$ for the values $N=2\cdots 5$, respectively, as $w_{T}\to \pm \infty$. Also a second phase transition can be realized at lower redshifts $z\sim 1.43$, $1.70$, $1.88$ and $1.95$ for the values $N=2\cdots 5$, respectively. The model forecasts a smooth crossing of the phantom divide line at future to quintessencelike regime. Remarkably, the phase transitions of torsion gravity is associated to crossing the matter density parameter, $\Omega_{m}(z)$, the unit boundary line as clear in Figs. \ref{Fig:Mod2}\subref{fig:Mod2-Om} and \ref{Fig:Mod2}\subref{fig:Mod2-wT}.

In the following part of this subsection, we show how the model parameter can be constrained to produce viable cosmic evolution. This means that if the predicted values of the model parameters agree with their measured ones, the assumed parametrization could be a good approximation to describe the cosmic history. As a matter of fact, we need the matter density parameter to reach a maximal value $\Omega_{m}(z)=1$ asymptotically, i.e as $z\to \infty$. This condition is useful to add a further constraint on the free parameters $q_0$ and $q_1$. In more detail, we write the asymptotic expansion of the matter density parameter (\ref{Mod2-Omega_m}) up to the second order of the redshift
$$\widetilde{\Omega}_{m}(z)\thickapprox \frac{\Omega_{m,0}}{N^{2r}}z^{1-2q_0+2q_1\ln{N}}.$$
For viable models, we need $\widetilde{\Omega}_{m}(z)\leq 1$, otherwise the torsion density parameter should drop to negative values. This constrains $q_{1}$ to a minimum value
\begin{equation}
\nonumber     q_{1}\leq \lim_{z\to \infty}-\frac{1}{2} \frac{\left( \ln{\frac{1}{\Omega_{m,0}z}} +2q_0\ln{z}\right)  \left( N-1 \right)}{\left( N-\ln{\frac{1}{z}} +N\ln{\frac{1}{z}} \right)\ln{N} }= \frac{2q_{0}-1}{2\ln{N}}.
\end{equation}
\begin{table*}[t!]
\caption{\label{Table2}%
The main results of model 2 according to the values of ($q_0,q_1$) parameters as given in Ref. \cite{Mamon:2016dlv} and by using the matter density parameter constraint (\ref{q2-min}). In the later, we keep the strict measured values $q_0$ as they are, while choosing the corresponding values of $q_1$ to fulfill the constraint.}
\begin{ruledtabular}
\begin{tabular*}{\textwidth}{lccccccc}
\textbf{Dataset}                        &$N$& $q_0$ & $q_1$ & $f(T)/\Lambda$CDM           & $\Omega_m(z)\leq 1$          & \textbf{Torsion} & \textbf{Viability} \\
                                        & &       &       & \textbf{compatibility} & \textbf{constraint} & \textbf{EoS}, $\omega_T$           &  \\
\colrule
JLA SNIa+BAO/CMB                       & 2 &  $-0.45$  &  $-2.56$  & not                 & violated                    & diverges\footnote{\label{footnote:2a}Note that the torsion EoS diverges twice as the matter density parameter crosses the unit boundary line twice, see Figs. \ref{Fig:Mod2}\subref{fig:Mod2-Om} and \ref{Fig:Mod2}\subref{fig:Mod2-wT}.}     & not        \\
JLA SNIa+BAO/CMB                       & 3 &  $-0.54$  &  $-1.35$  & not                 & violated                    & diverges$^\textrm{\ref{footnote:2a}}$     & not        \\
JLA SNIa+BAO/CMB                       & 4 &  $-0.56$  &  $-1.03$  & not                 & violated                    & diverges$^\textrm{\ref{footnote:2a}}$     & not        \\
JLA SNIa+BAO/CMB                       & 5 &  $-0.56$  &  $-0.85$  & not                 & violated                    & diverges$^\textrm{\ref{footnote:2a}}$     & not        \\
\colrule
\multicolumn{8}{l}{\textbf{Using constraint (\ref{q2-min})}}\\
\colrule
JLA SNIa+BAO/CMB                       & 2 &  $-0.45$  &  $-3.366$  & not                 & fulfilled                   & does not diverge & not\footnote{\label{footnote:2b}Although the enhanced parameters-using the matter density parameter constraint (\ref{q2-min})- give smooth $\omega_T$ patterns (see Fig. \ref{Fig:Mod2A}\subref{fig:Mod2A-wT}), it cannot produce $f(T)$ gravity with better compatibility with $\Lambda$CDM as seen in Fig. \ref{Fig:Mod2A}\subref{fig:Mod2A-fz}. }       \\
JLA SNIa+BAO/CMB                       & 3 &  $-0.54$  &  $-1.525$  & not                 & fulfilled                   & does not diverge & not$^\textrm{\ref{footnote:2b}}$       \\
JLA SNIa+BAO/CMB                       & 4 &  $-0.56$  &  $-1.114$  & not                 & fulfilled                   & does not diverge & not$^\textrm{\ref{footnote:2b}}$       \\
JLA SNIa+BAO/CMB                       & 5 &  $-0.56$  &  $-0.913$  & not                 & fulfilled                   & does not diverge & not$^\textrm{\ref{footnote:2b}}$
\end{tabular*}
\end{ruledtabular}
\end{table*}

Indeed this constraint derives the parametrization (\ref{Mod2-Xz}) to perform $q\to 1/2$ as $z\to \infty$ just as mentioned in \cite{Mamon:2016dlv}. However, in practice this will not prevent the matter density parameter to cross the unit boundary line at smaller redshifts, it just flattens the peaks over a very wide redshift range. In this sense, one may rather need to control the $\Omega_{m}(z)$-peak amplitude to not cross the unit boundary line. In order to make this model viable at least at low redshifts $z=z_l$, we use the constraint $\Omega_m(z_l)=1$. Using (\ref{Mod2-Omega_m}), we solve this constraint for $q_1$,
\begin{equation}\label{q2-min}
    q_1=\frac{(N-1)(1+z_l)\left[\ln{(1+z_l)^{2q_0-1}}-\ln{\Omega_{m,0}}\right]}{\ln\left[\frac{(1+z_l)^{2\left((N-1)\ln{N}-1\right)(1+z_l)}\left(N+z_l\right)^{2(N+z_l)}}{N^{2N(1+z_l)}}\right]}.
\end{equation}
It is reasonable to keep $q_0$ as measured in \cite{Mamon:2016dlv} assuming the $\Omega_m(z_l)$-peak to occur at $z_l=3$, so we get $q_1\sim -3.366$, $-1.525$, $-1.114$ and $-0.913$ for the values $N=2\cdots 5$, respectively. In this case, we plot the evolution $f(T)$ gravity, namely (\ref{Mod2-f(z)}), as shown in Fig. \ref{Fig:Mod2A}\subref{fig:Mod2A-fz}. By comparison with Fig. \ref{Fig:Mod2}\subref{fig:Mod2-fz}, we find that the plots still do not match the $\Lambda$CDM model at large redshifts. On the other hand, the universe effectively cannot describe sCDM matter domination era correctly as $w_{eff}> 0$ at large redshifts, see Fig. \ref{Fig:Mod2A}\subref{fig:Mod2A-weff}. Indeed, the matter density parameter does not cross the unit boundary line, but the behavior still inconsistent with the matter domination era. This can be shown clearly in Fig. \ref{Fig:Mod2A}\subref{fig:Mod2A-Om}. Finally, we note that the torsional EoS in the enhanced version does not indicate phase transitions anymore, whereas $w_T$ becomes finite at all redshifts, see Fig. \ref{Fig:Mod2A}\subref{fig:Mod2A-wT}. In Table \ref{Table2}, we summarize the results of model 1 according to the values of the model parameters $q_0$ and $q_1$ as given in Ref.  \cite{Mamon:2016dlv} and by applying the matter density constraint (\ref{q2-min}).

In conclusion, the model parameters in the enhanced version neither match the best fit values as measured by different datasets combinations nor produce a viable behavior of the matter density parameter. Therefore, it cannot be considered as a viable model to produce the whole cosmic history.
\subsection{Model 3}\label{Sec4.4}
\begin{figure*}[t!]
\centering
\subfigure[~$w_{eff}$ according to \cite{Mukherjee:2016eqj}]
{\label{fig:Mod3-weff}\includegraphics[scale=.26]{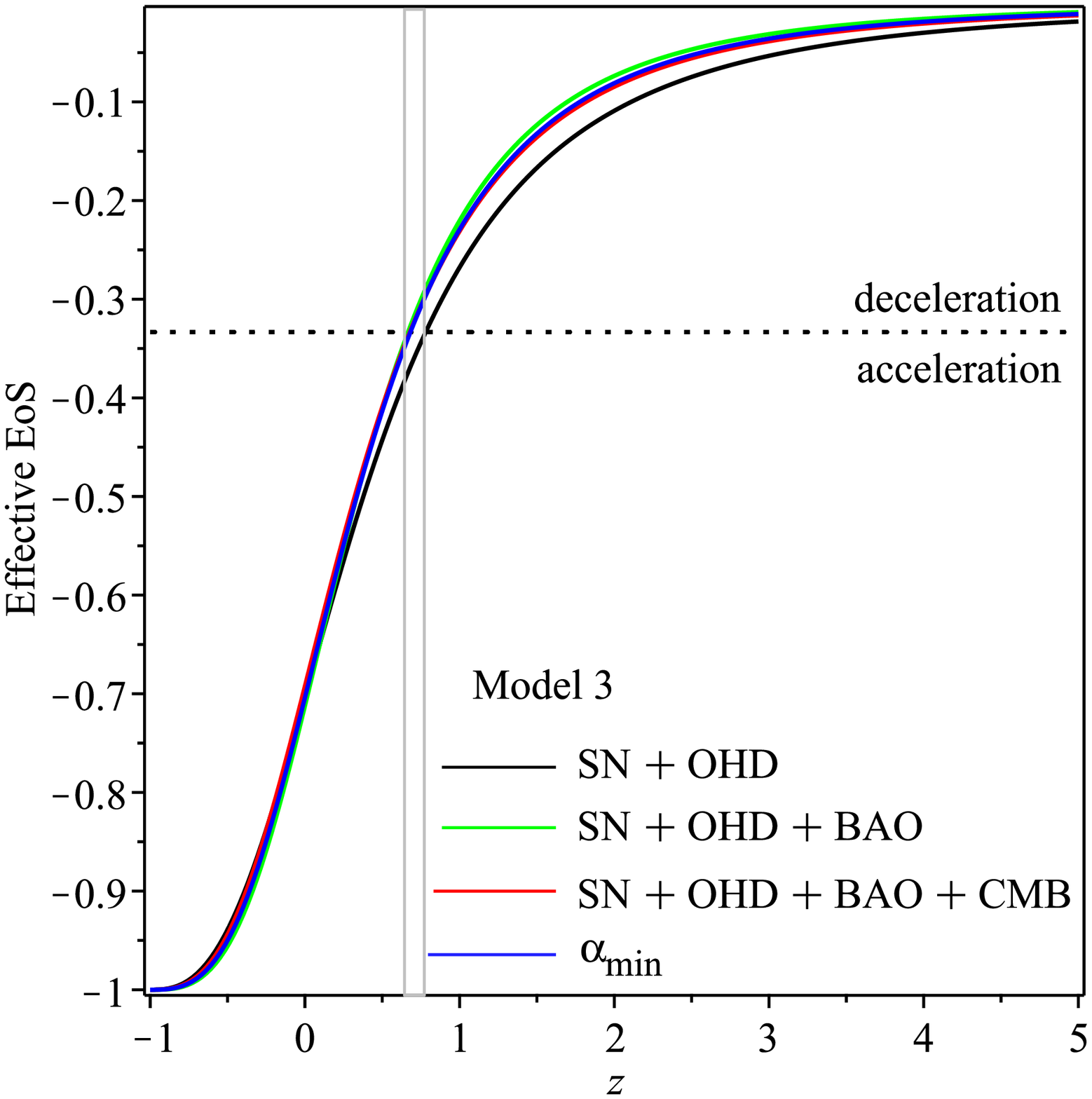}}
\subfigure[~$\Omega_{m}(z)$, $\Omega_{T}(z)$ according to \cite{Mukherjee:2016eqj}]
{\label{fig:Mod3-Om}\includegraphics[scale=.26]{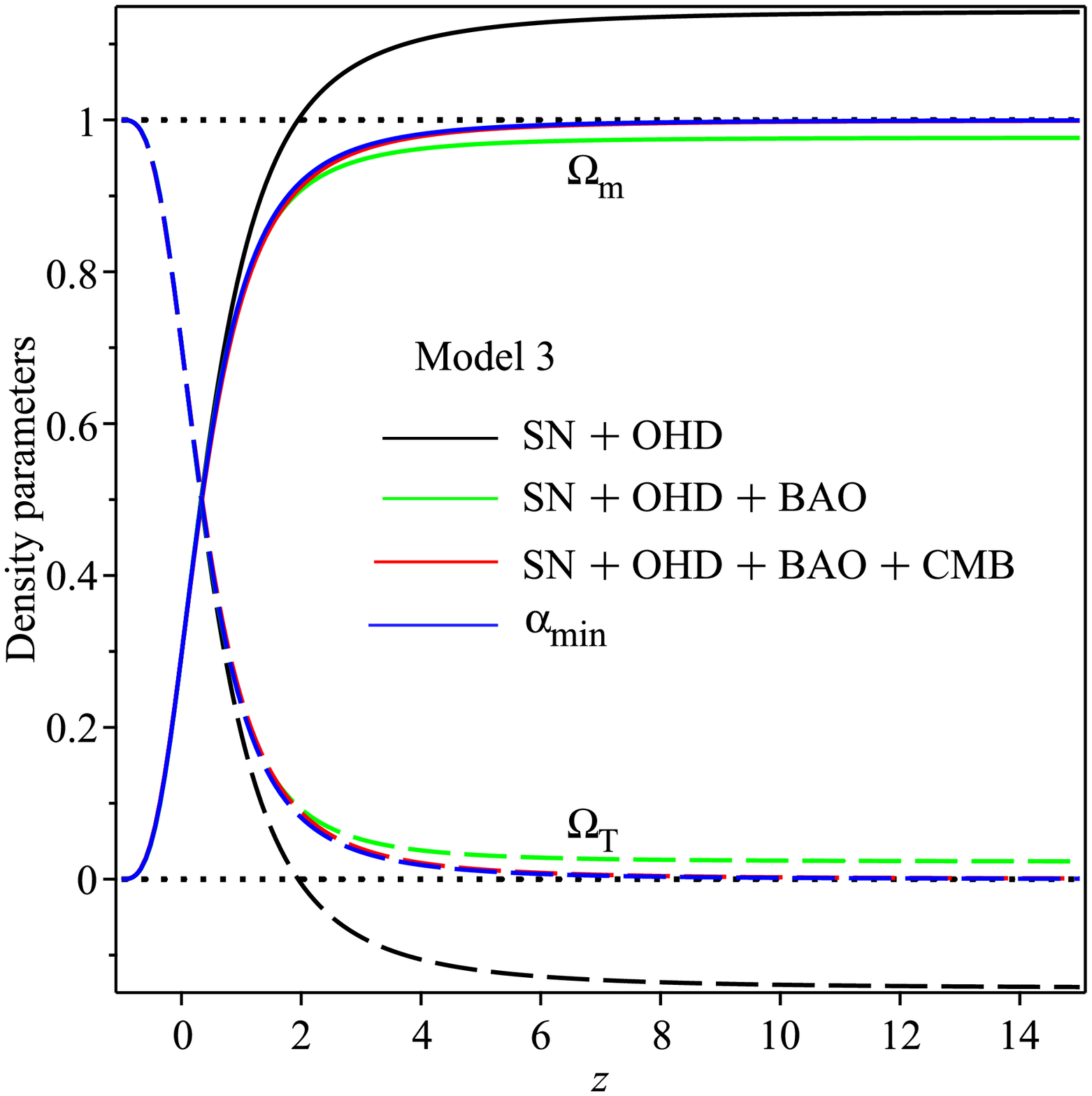}}
\subfigure[~$w_{T}$ according to \cite{Mukherjee:2016eqj}]
{\label{fig:Mod3-wT}\includegraphics[scale=.26]{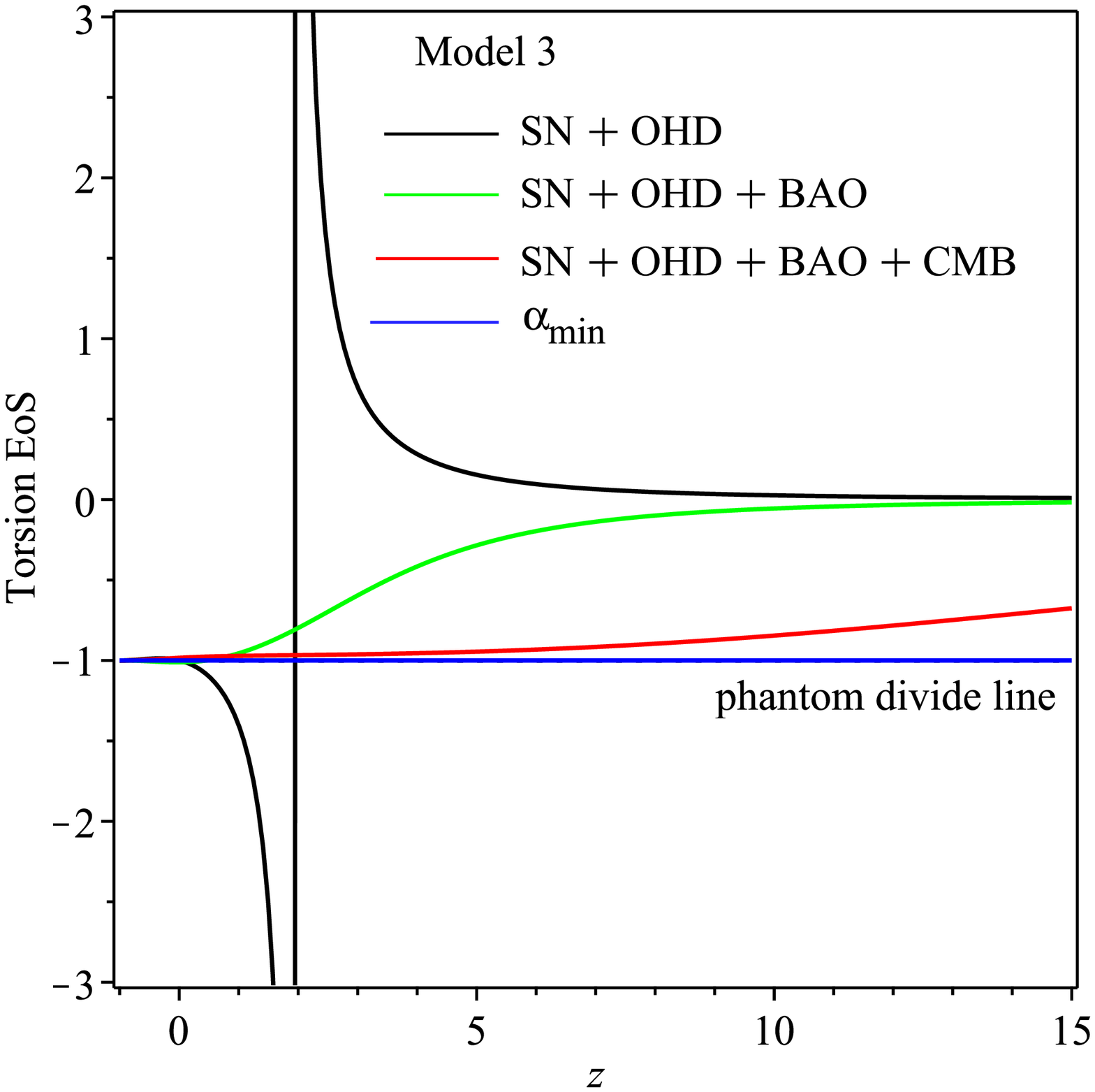}}
\subfigure[~$w_{eff}$ using the constraint (\ref{alpha-min})]
{\label{fig:Mod3A-weff}\includegraphics[scale=.26]{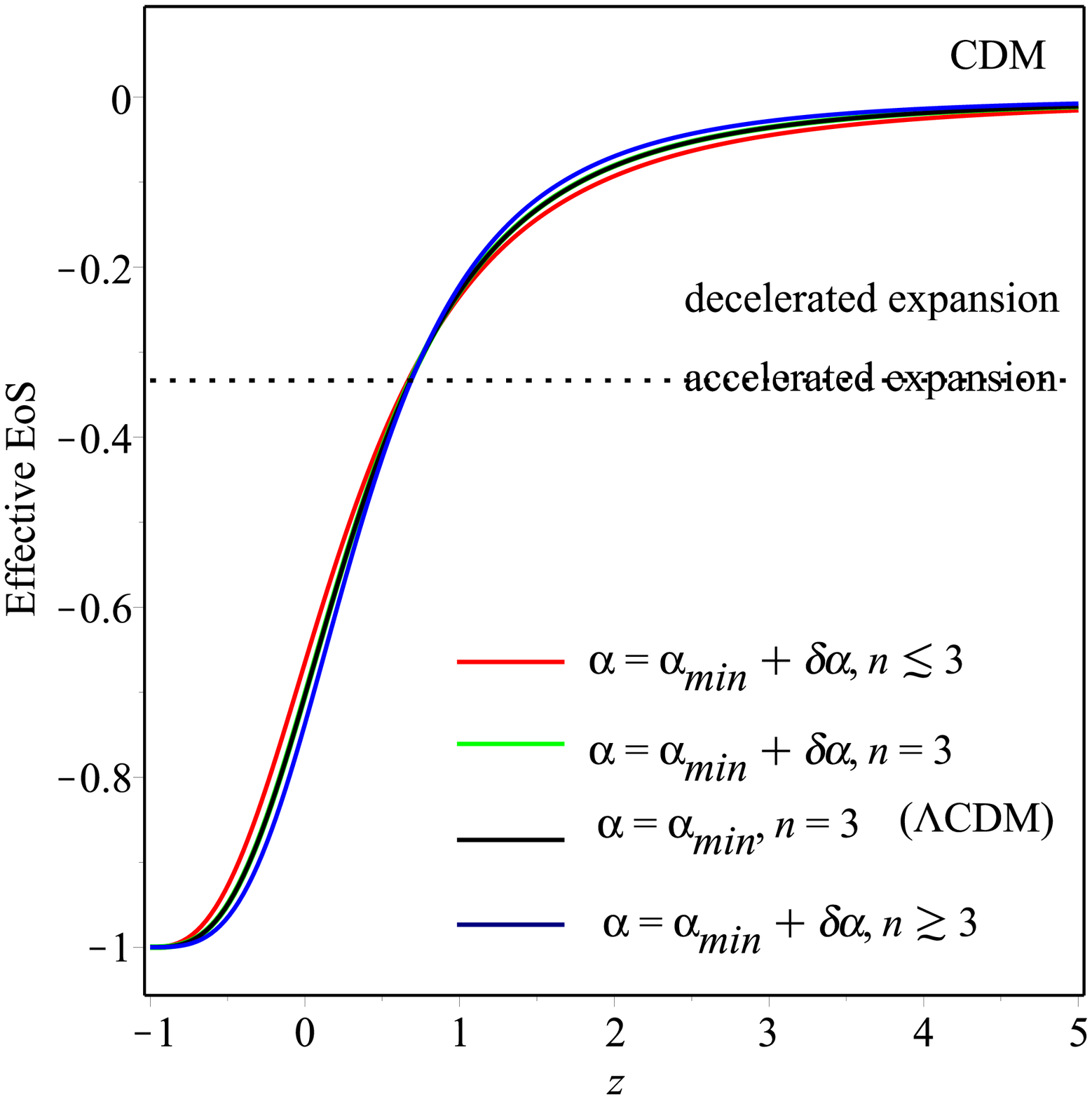}}
\subfigure[~$\Omega_{m}$, $\Omega_{T}$ using the constraint (\ref{alpha-min})]
{\label{fig:Mod3A-Om}\includegraphics[scale=.26]{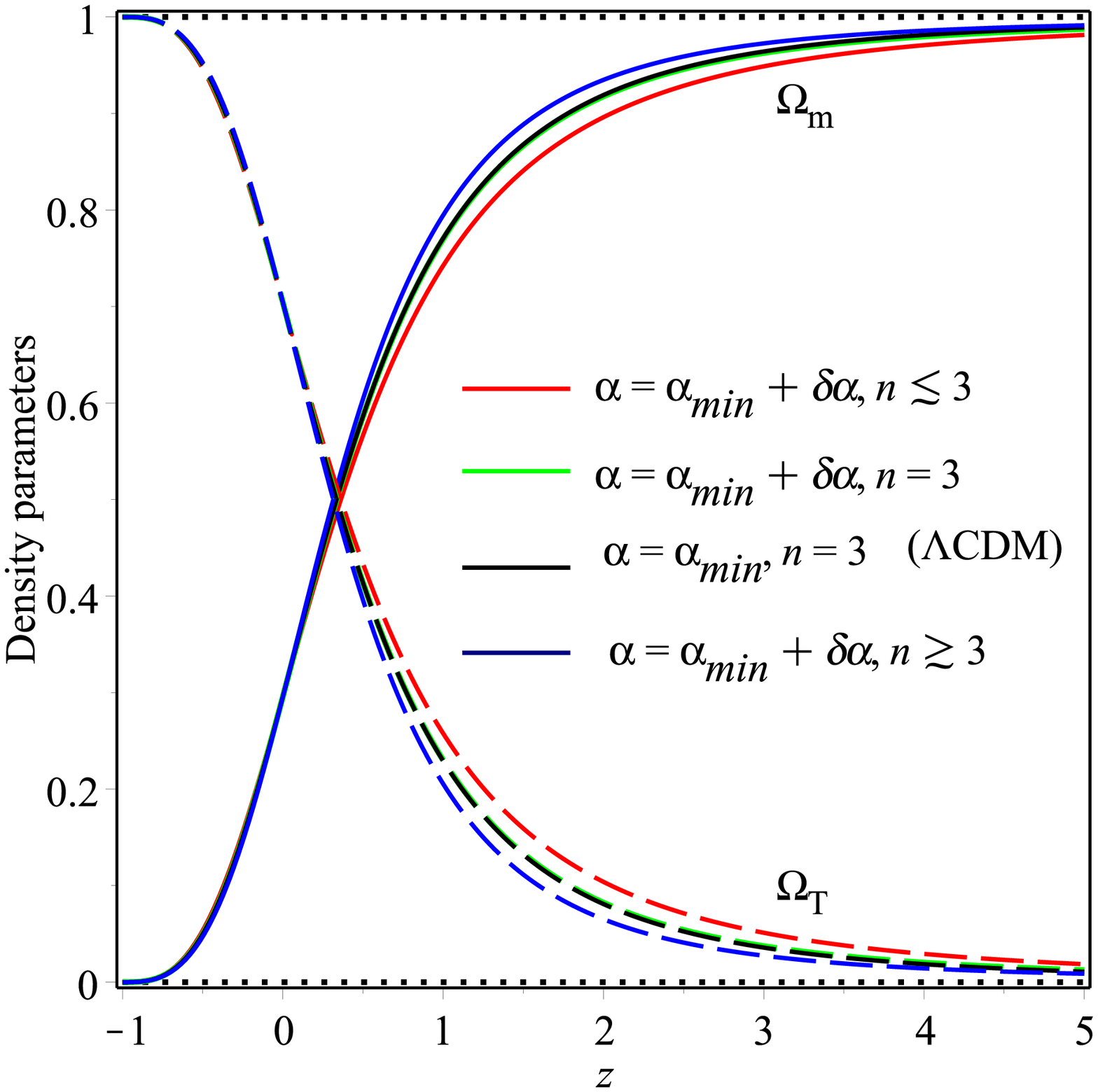}}
\subfigure[~$w_{T}$ using the constraint (\ref{alpha-min})]
{\label{fig:Mod3A-wT}\includegraphics[scale=.26]{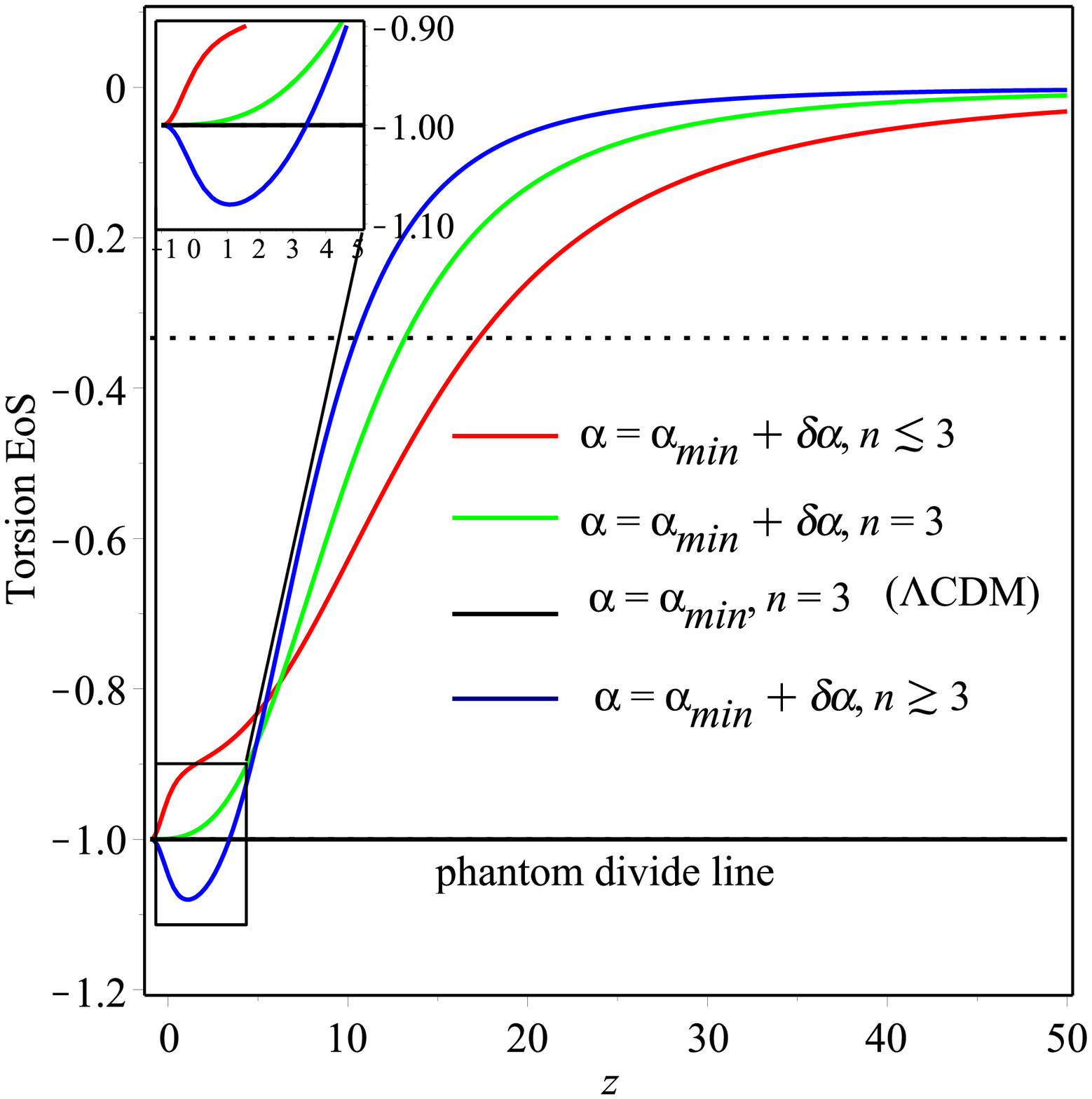}}
\caption[figtopcap]{In subfigs. \subref{fig:Mod3-weff} -- \subref{fig:Mod3-wT}, the best fit values of the model parameters ($\alpha$, $n$) are taken according to the dataset combination as measured in \cite{Mukherjee:2016eqj}; (0.445, 2.8) for SN+OHD,
(0.409, 3.13) for SN+OHD+BAO, (0.444, 2.907) for SN+OHD+BAO+CMB and ($\alpha = \frac{\Omega_{m,0}}{1-\Omega_{m,0}}$, $n$ = 3) for $\Lambda$CDM with Planck parameters.
In subfigs. \subref{fig:Mod3A-weff} -- \subref{fig:Mod3A-wT}, we use the constraint (\ref{alpha-min}) to determine $\alpha_{min}$ for different choices of $n$, while the additive constant is taken as $\delta \alpha=10^{-3}$.}
\label{Fig:Mod3}
\end{figure*}

In this subsection, we reconstruct the $f(T)$ gravity theory upon the parametric form of the effective EoS assumed in Ref.~\cite{Mukherjee:2016eqj},
\begin{equation}\label{Mod3-weff}
    w_{eff}=-\frac{1}{1+\alpha(1+z)^{n}},
\end{equation}
\begin{figure}[b!]
\centering
\includegraphics[scale=0.4]{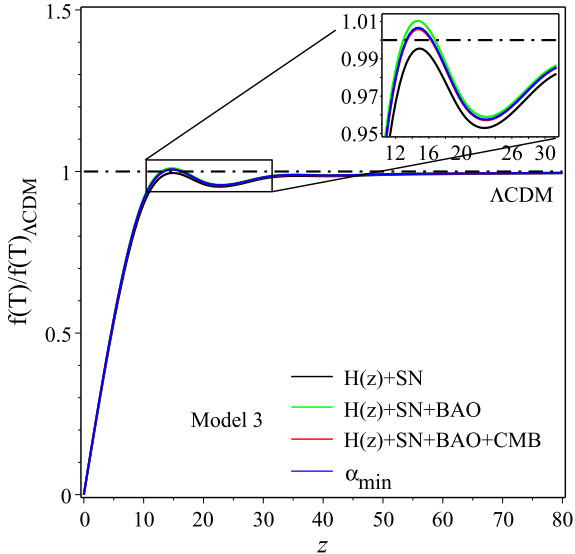}
\caption{Evolution of $f(T(z))$ gravity, (\ref{Mod3-f(z)}), vs $\Lambda$CDM, (\ref{fLCDM}): The model parameters ($\alpha$, $n$) are taken from \cite{Mukherjee:2016eqj} according to the datasets combinations: SN+OHD (0.445, 2.8), SN+OHD+BAO (0.409, 3.13) and SN+OHD+BAO+CMB (0.444, 2.907), while the $\alpha_{min}$ is taken for the value $n=3$ with the constraint (\ref{alpha-min}).}
\label{Fig:Mod3-fz}
\end{figure}
where $\alpha$ and $n$ are two model parameters. At large redshifts, the universe effectively produces sCDM as $w_{eff}\to 0$. At far future $z\to -1$, the universe effectively evolves toward de Sitter as $w_{eff}\to -1$. This can be shown clearly in Fig. \ref{Fig:Mod3}\subref{fig:Mod3-weff}. In principal, this pattern can produce a successful cosmic history. Using (\ref{deceleration}), the deceleration parameter associated with the above parametrization can be written as
\begin{equation}\label{Mod3-qz}
   q(z)=-1+\frac{3\alpha(1+z)^{n}}{2\left[1+\alpha(1+z)^{n}\right]}.
\end{equation}
Inserting the above into (\ref{Hubble-deceleration}), the Hubble parameter reads
\begin{equation}\label{Mod3-H}
    H(z)=H_{0}\left[\frac{1+\alpha(1+z)^{n}}{1+\alpha}\right]^{\frac{3}{2n}}.
\end{equation}
One can see that the model produces $\Lambda$CDM model as a particular case when $n=3$, then we obtain that $\Omega_{m,0}=\frac{\alpha}{1+\alpha}$ or equivalently $\alpha=\frac{\Omega_{m,0}}{1-\Omega_{m,0}}$. We take the value $\Omega_{m,0}=0.297$ as measured in Ref. \cite{Mukherjee:2016eqj}. In viable dynamical DE models, one may expect that the value of $n$ to be close to $n=3$. On the other hand, we use the second reconstruction equation (\ref{Reconstruction2}) to evaluate the $f(T)$ gravity which generates the parametric form (\ref{Mod3-weff}). We obtain
\begin{eqnarray}
\nonumber    f(z)&=&-9\alpha\Omega_{m,0}H_0^2\left[1+\alpha(1+z)^{n}\right]^{\frac{3}{2n}}\\
&&\times    \int_0^z{\frac{(1+\bar{z})^{n+2}}{\left[1+\alpha(1+\bar{z})^{n}\right]^{1+\frac{3}{2n}}}}d\bar{z}.\label{Mod3-f(z)}
\end{eqnarray}
In Fig. \ref{Fig:Mod3-fz}, we plot the evolution of the $f(T)$ gravity at hand versus $\Lambda$CDM for different values of the model parameters $\alpha$ and $n$ according to the dataset used \cite{Mukherjee:2016eqj}. The plots show systematic deviation of the $f(T)$ theory from $\Lambda$CDM when the dataset combination SN+observed Hubble data (OHD) is used, since it does not oscillate about $\Lambda$CDM. Otherwise, the theory is compatible with $\Lambda$CDM. We will give the reasons for these results later in this subsection.

Using the deceleration (\ref{Mod3-qz}), the matter density parameter (\ref{matter-density-parameter}) reads
\begin{equation}\label{Mod3-Omega_m}
    \Omega_{m}(z)=\Omega_{m,0}\,(1+z)^{3}\left[\frac{1+\alpha}{1+\alpha(1+z)^{n}}\right]^{\frac{3}{n}}.
\end{equation}
In Fig. \ref{Fig:Mod3}\subref{fig:Mod3-Om}, we plot the evolution of the matter density parameter using different values of the model parameters. As seen, the density matter exceeds the unity at redshift $z \sim 2$ when the parameters are fitted with the dataset SN+OHD, on the other hand the density matter has a slight but not trivial deviation form $\Lambda$CDM at large $z$ when the parameters are fitted with the dataset SN+OHD+BAO. However, it evolves very similar to $\Lambda$CDM when the CMB (shift parameter) is added. In fact, these results are in agreement with the plots of Fig. \ref{Fig:Mod3-fz}.

Substituting from (\ref{Mod3-H}) and (\ref{Mod3-f(z)}) in (\ref{wT(z)}), we evaluate the torsion (DE) EoS
\begin{equation}\label{Mod3-wDE}
    w_{T}=-\frac{\left[1+\alpha(1+z)^n\right]^{\frac{3}{n}}-\alpha(1+z)^n \left[1+\alpha(1+z)^n\right]^{\frac{3}{n}-1}}{\left[1+\alpha(1+z)^n\right]^{\frac{3}{n}}-\Omega_{m,0}(1+z)^3 (1+\alpha)^{\frac{3}{n}}}.
\end{equation}
As seen from Fig. \ref{Fig:Mod3}\subref{fig:Mod3-wT} that the torsional EoS diverges at redshift $z\sim 2$ when the dataset SN+OHD is used. We note that the torsional phase transition is associated with the crossing of the matter density parameter of the unit boundary line as seen in Fig \ref{Fig:Mod3}\subref{fig:Mod3-Om}.
\begin{table*}[t!]
\caption{\label{Table3}%
The main results of model 3 according to the values of ($\alpha,~n$) parameters as given in Ref. \cite{Mukherjee:2016eqj} and by using the matter density parameter constraint (\ref{alpha-min}), where $\alpha=\alpha_{min}+\delta \alpha (=10^{-3})$. In all treatments we take $\Omega_{m,0}=0.297$ as derived in \cite{Mukherjee:2016eqj}.}
\begin{ruledtabular}
\begin{tabular}{lcccccc}
\textbf{Dataset}            & $\alpha$ & $n$ & $f(T)/\Lambda$CDM           & $\Omega_m(z)\leq 1$          & \textbf{Torsion} & \textbf{Viability} \\
                            &          &       & \textbf{compatibility} & \textbf{constraint} & \textbf{EoS}, $\omega_T$           &  \\
\colrule
SN+OHD                      &   $0.445$\footnote{\label{footnote:3a}Note that this dataset provides $\alpha < \alpha_{min}$, which explains the violation of the matter density constraint.}   & $2.8$  & not                & violated                  & diverges     & not        \\
SN+OHD+BAO                  &   $0.409$    & $3.13$  & semi               & fulfilled                 & does not diverge     & yes        \\
SN+OHD+BAO+CMB              &   $0.444$    & $2.907$  & semi             & fulfilled                 & does not diverge     & yes        \\
\colrule
\multicolumn{7}{l}{\textbf{Using constraint (\ref{alpha-min})}}\\
\colrule
case (i)                      &   $0.503$    & $2.7 \lesssim 3$  & semi               & fulfilled                   & does not diverge (quintessence) & yes      \\
case (ii)                  &   $0.356$    & $3.3 \gtrsim3$  & semi               & fulfilled                   & does not diverge (quintom) & yes      \\
case (iii)              &   $0.421$    & $3$  & semi               & fulfilled                   & does not diverge (quintessence) & yes      \\
\colrule
$\Lambda$CDM                            & $\alpha_{min}$   & 3 & yes                & fulfilled                   & $-1$           & yes
\end{tabular}
\end{ruledtabular}
\end{table*}

In order to constrain the model parameters, we follow the treatment of Sec. \ref{Sec4.1} by requiring the matter density parameter (\ref{Mod3-Omega_m}) to reach a maximal value $\Omega_{m,0}=1$ asymptotically, i.e as $z\to \infty$. This condition is useful to put a lower bound on the parameter $\alpha$. In more detail, we write the leading term of the asymptotic expansion of the matter density parameter
$$\widetilde{\Omega}_{m}(z)\thickapprox \Omega_{m,0} \left(1+\frac{1}{\alpha}\right)^{\frac{3}{n}}.$$
For viable models, we need $\widetilde{\Omega}_{m}(z)\leq 1$, otherwise the torsion density parameter would drop below zero. This constrains $\alpha$ to a minimum value
\begin{equation}\label{alpha-min}
   \alpha_{min}=\frac{\Omega_{m,0}^{\frac{n}{3}}}{1-\Omega_{m,0}^{\frac{n}{3}}}.
\end{equation}
If $\alpha$ goes below the above minimum, the matter density parameter would exceed the unity at some redshift at past, and subsequently the torsion density parameter becomes negative. For example, the measured value of the parameter $\alpha=0.445$ according to the dataset SN+OHD \cite{Mukherjee:2016eqj} is less than the allowed minimum value $\alpha_{min}=0.4728$ according to the density matter constraint\footnote{Note that we take $\Omega_{m,0}=0.297$ as derived in \cite{Mukherjee:2016eqj} and $n=2.8$ as measured by using the dataset SN+OHD.} (\ref{alpha-min}). Therefore, we understand the incompatibility of the model results (see Fig. \ref{Fig:Mod3}\subref{fig:Mod3-Om}) whenever the dataset SN+OHD is used. On the contrary, we find that the measured values $\alpha=0.409 > \alpha_{min}=0.3904$ (using SN+OHD+BAO dataset) and $\alpha=0.444 > \alpha_{min}=0.4438$ (using SN+OHD+BAO dataset) are compatible and give viable cosmic scenarios.

Notably, for the $\Lambda$CDM case ($n=3$), the minimum value $\alpha_{min}=\frac{\Omega_{m,0}}{\Omega_{\Lambda,0}}$ represents the ratio between the matter and the torsion density parameters at present. In Figs \ref{Fig:Mod3}\subref{fig:Mod3A-weff}--\subref{fig:Mod3A-wT}, we plot $\Lambda$CDM as $n=3$ and $\alpha=\alpha_{min}$, at those values we obtain a fixed torsional EoS $w_T=-1$. If $\alpha$ slightly exceeds $\alpha_{min}$, we list three possible viable cases:
(i) For $n\lesssim 3$, we find that $w_T$ evolves in quintessence with $w_T \gtrsim -1$ at present.
(ii) For $n\gtrsim 3$, the torsional EoS has a quintom behavior, since $w_T$ crosses the phantom divide line at $z\sim 4$ from quintessence to phantom with $w_T \lesssim -1$ at present.
(iii) For $n=3$, the torsional EoS evolves in a quintessence regime with $w_T\sim -1$ at present.  In all cases, $w_T\to 0$ at large redshifts, which explains the late accelerating expansion, and evolves toward a cosmological constant (pure de Sitter) at future. We summarize the model results in Table \ref{Table3}.

In conclusion, we find that the effective EoS parametrization (\ref{Mod3-weff}) produces a viable $f(T)$ gravity model. However, the model parameters $n$ and $\alpha$ can be better constrained by the upcoming DE surveys to determine the nature of the DE precisely. On the other hand, the $f(T)$ theory presented in this subsection is capable to produce a dynamical torsional DE, then--in principal--it could be useful to reconcile the local measurement of the Hubble constant with its global measured value. More interestingly the $f(T)$ theory at hand is ready to be tested on the perturbation level too.
\section{Summary}\label{Sec5}
Recent attempts to describe the late accelerating expansion of the universe, via kinematic approach by considering some parametric forms of the deceleration parameter, have been discussed. Although some of these parametric forms could be useful to encode the deceleration-to-acceleration transition, those need to be treated within a dynamical framework or modified gravity. This allows not only for more tests of other cosmological parameters on the background level but also for further tests on the perturbation level of the theory. In this paper, we have set a reconstruction method of $f(T)$ gravity which generates any particular $q(z)$--parametrization. This has been achieved by recognizing the compatibility of the deceleration parameter (\ref{Hubble-deceleration}) and $f(T)$ gravity (\ref{phase-portrait-z}). In addition, we have derived two more reconstruction equations by knowing the effective EoS $w_{eff}(z)$ or the DE EoS $w_{DE}(z)$.

We have examined three models in this paper:
For model 1, the $q(z)$ parametrization (\ref{Mod1-Xz}) has been adopted. We compare the corresponding $f(T)$ gravity with $\Lambda$CDM model showing the inviability of the model even by enhancing its model parameters.
For model 2, the $q(z)$ parametrization (\ref{Mod2-Xz}) has been adopted. Similar to model 1, it cannot produce a viable cosmological scenario.
For model 3, the $w_{eff}(z)$ parametrization (\ref{Mod3-weff}) has been adopted. The corresponding $f(T)$ gravity shows a good compatibility with $\Lambda$CDM results as well as the current observations. Since the model is flexible to produce a dynamical DE model with quintessence or quintom behavior, we expect the corresponding $f(T)$ gravity to explain the nature of the DE beyond $\Lambda$CDM.

In these three models, the torsional EoS diverges if the matter density parameter crosses the unit boundary line (alternatively the torsion density parameter becomes negative). This feature can be verified by the upcoming DE surveys.

In this paper, we have examined the reconstruction method on the background level of the obtained theories. However, it is more interesting to examine the theory on the perturbation level as well. We leave this task for future work.


\begin{acknowledgments}
We would like to thank Amr El-Zant for helpful discussions.
\end{acknowledgments}

%

\end{document}